\documentclass[a4paper,fleqn,usenatbib]{mnras}
\usepackage{newtxtext,newtxmath}
\usepackage{hyperref}
\usepackage[T1]{fontenc}
\usepackage{epsfig}
\usepackage{ae,aecompl}
\usepackage{graphicx}    
\usepackage{amsmath}     
\usepackage{gensymb}     
\usepackage{pdflscape}
\usepackage{comment}
\usepackage{subcaption}
\usepackage{breakurl}

\hypersetup{pdftitle=On the origin of soft-excess of Ark 120}

\title[Origin of soft-excess of Ark 120]{Long term X-Ray Observations of Seyfert 1 Galaxy Ark 120: On the origin of soft-excess}

\author[P Nandi et al.]{
Prantik Nandi$^{1}$\thanks{E-mail: prantiknandi@bose.res.in},
Arka Chatterjee$^{1}$\thanks{E-mail: arkachatterjee@bose.res.in},
Sandip K. Chakrabarti$^{2}$\thanks{E-mail: sandip@csp.res.in},
Broja G. Dutta$^{2,3}$\thanks{E-mail: brojadutta@gmail.com}
\\
$^{1}$Department of Astrophysics \& Cosmology, S. N. Bose National Centre for Basic Science, Salt lake, Sector III, Kolkata 700091, India\\
$^{2}$Indian Centre for Space Science, Garia Station Road, Kolkata 700084, India \\
$^{3}$Department of Physics, Rishi Bankim Chandra College, Naihati, West Bengal, 743165, India
}

\date{Accepted XXX. Received YYY; in original form ZZZ}

\pubyear{2020}

\begin{document}
\label{firstpage}
\pagerange{\pageref{firstpage}--\pageref{lastpage}}
\maketitle

\begin{abstract}
We present the long-term X-ray spectral and temporal analysis of a `bare-type AGN'
Ark 120. We consider the observations from {\it XMM-Newton}, {\it Suzaku}, 
{\it Swift}, and {\it NuSTAR} from 2003 to 2018. The spectral properties of this 
source are studied using various phenomenological and physical models present in the literature. We report 
(a) the variations of several physical parameters, such as the temperature and optical depth 
of the electron cloud, the size of the Compton cloud, and accretion rate 
for the last fifteen years. The spectral variations are explained from the change in the 
accretion dynamics; (b) the X-ray time delay between 0.2-2 keV and 
3-10 keV light-curves exhibited {\it zero-delay} in 2003, {\it positive delay} of $4.71\pm2.1$ 
ks in 2013, and {\it negative delay} of $4.15\pm1.5$ ks in 2014. The delays are 
explained considering Comptonization, reflection, and light-crossing time; 
(c) the long term intrinsic luminosities, obtained using {\tt nthcomp}, of the soft-excess 
and the primary continuum show a correlation with a Pearson Correlation Co-efficient 
of $0.922$. This indicates that the soft-excess and the primary continuum are originated 
from the same physical process. From a physical model fitting, we infer that the soft excess 
for Ark 120 could be due to a small number of scatterings in 
the Compton cloud. Using Monte-Carlo simulations, we show that indeed
the spectra corresponding to fewer scatterings could provide a steeper 
soft-excess power-law in the 0.2-3 keV range. Simulated luminosities 
are found to be in agreement with the observed values. 

\end{abstract}




\section{Introduction}
\label{sec:intro}
Active Galactic Nuclei (AGNs) are the most energetic phenomena in the universe. The 
emitted radiation is observed over the entire range of the electromagnetic spectrum. 
The high energy X-rays are believed to be emitted from the innermost 
region of an accretion disc which surrounds the central black hole \citep{SS1973, PRP73}. 
The X-ray spectra of Seyfert 1 galaxies, a subclass of AGNs, is mostly fitted 
by a power-law component with photon index 
in the range $\Gamma=1.6-2.2$ \citep{Bianchi2009, Sobolewska2009} and a high energy cut-off. 
The spectral contribution which deviates from the power-law at lower energy (below $\sim 2$ keV) is known 
as `soft excess' \citep{Halpern84, Arnaud1985, Singh1985}. The X-ray spectra are often associated with 
a Fe K$\alpha$ line, which is observed near 6.4 keV, and a Compton hump in the energy range of 20.0 to 40.0 keV. 
It has been observed that the primary power-law emission is produced by the Comptonization of 
low energy seed photons \citep{Sunyaev1980, Titarchuk1994} emitted from the standard Keplerian disc. 
The seed photons are processed from the accretion mechanism, and the peak emission arises at optical/ultraviolet 
(UV) wavelengths \citep{PRP73} for a supermassive black hole (SMBH). However, the location, as 
well as the geometry of the Compton reprocessing region, are still a matter of debate. This Compton cloud 
can be situated above the accretion disc \citep{Haardt1991, Haardt1993, Poutanen1996} or at the base of 
the relativistic jet \citep{CT95,Fender1999, Fender2004, Markoff2005}. The region could be a hot, 
radiatively inefficient and behave like a quasi-Bondi flow
as discussed initially by \cite{Ichimaru1977}. This region could originate 
the thermal Comptonization of soft photons produced in the optical/UV range 
from an optically thick Keplerian disc \citep{Magdziarz1998, Dewangan2007, Done2012, Lohfink2012} or a blurred reflection 
from ionized disc \citep{Fabian2002, Ross2005, Crummy2006, Garcia2014}. The iron line 
is thought to be originated by the photoelectric absorption followed by the fluorescence line 
emission from a dense and relatively cold accretion disc. Moreover, it is believed 
that the Compton hump could be due to the Compton scattering dominated above 10 keV in a 
relatively cold dense medium. Nevertheless, the complex broad-band spectrum of AGNs requires a
proper physical explanation of the flow dynamics and radiative properties around 
the central engine across the soft and hard energy regime of the X-ray.   
      
In this scenario, the Two-Component Advective Flow (TCAF) \citep{CT95} model, 
which combines the essence of all the salient features of a viscous transonic flow  
\citep{Chakrabarti1989, Chakrabarti1990, Chakrabarti1995} around black holes is worth exploring. It is a
physical solution encompassing hydrodynamics and radiative processes. The transonic 
flow solution allows two types of accretion flows depending on how efficiently angular 
momentum is being transported: a viscous, geometrically thin, optically thick standard 
Keplerian component \citep{SS1973} and a weakly viscous, geometrically thick, optically thin 
sub-Keplerian halo component \citep{CT95}. The latter is basically an inefficiently radiating 
generalized Bondi flow with high radial velocity till it forms the centrifugal barrier after which it
becomes efficient in radiating at higher energies. The Keplerian disc is formally truncated at 
the centrifugal barrier, the outer boundary of which is the shock location \citep{Chakrabarti1989}. 
The post-shock region (i.e., the region between the shock and the innermost sonic point) is known as 
CENtrifugal barrier supported BOundary Layer or CENBOL and it acts as the Compton cloud. 
The soft photons from the Keplerian disc are upscattered by Comptonization process in the post-shock 
region and produce the high energy X-ray photons. TCAF, a self-consistent model, is quantified 
by four flow parameters: 
two types of accretion rates, namely, the disc rate ($\dot{m}_d$) and halo rate ($\dot{m}_h$), size and density
of the Compton cloud, through the shock location ($X_s$) and the compression ratio ($R$), ratio of the post-shock and the pre-shock flow densities ($\frac{\rho_{+}}{\rho_{-}}$).
It also requires an intrinsic parameter, namely, the mass of the central black hole (in the 
units of $M_{\odot}$), and an extrinsic parameter, namely, the normalization which is required to 
place the observed spectrum over the theoretical spectrum of TCAF. The broadband spectra 
of M87 was explained with this model by fitting the data from multi-wavelength 
observations\citep{MC2008}. Later, TCAF has been implemented in the {\tt xspec} as 
a local table model and has been successful to fit the data of the Galactic black holes 
\citep{DD2014} and has also been able to estimate the mass of nearby Seyfert 1 galaxy 
NGC 4151 using {\it NuSTAR} data \citep{Nandi2019}.

Arakelian 120 (Ark 120) is a nearby ($z=0.03271$\footnote{The redshift is taken from 
the NASA/Infrared Process and Analysis center (IPAC) Extragalactic Database.  
\burl{https://ned.ipac.caltech.edu}}) radio-quiet Seyfert 1 AGN with radio-loudness 
$R\approx0.1$ \citep{Condon1998, Ho2002}. This source was intensely monitored nearly 
in all wavelengths: optical/UV \citep{Kollatschny1981, Kollatschny1981b, Schulz1981, 
Alloin1988, Marziani1992, Peterson1998, Stanic2000, Popovic2001, Doroshenko2008, Kuehn2008} 
and X-ray \citep{Vaughan2004, Nardini2016, Reeves2016, Gliozzi2017, Lobban2018} and was
found to be consistently bright in optical, UV, and X-rays displaying substantial 
wavelength-dependent variability \citep{Gliozzi2017, Lobban2018}. From the simultaneous UV/X-ray 
measurements, it was reported that the observations are neither `contaminated' by absorption 
signatures along the line of sight \citep{Vaughan2004, Reeves2016, Crenshaw1999} nor by 
neutral intrinsic absorbers \citep{Reeves2016} around the central engine. Furthermore, 
Ark 120 is nearly free from intrinsic reddening in the IR-optical-UV continuum \citep{Ward1987, 
Vasudevan2009}. Therefore, it provides one of the cleanest views ($N_H\sim3\times10^{19}$ cm$^{-2}$; 
\citep{Vaughan2004}) of the central region. This type of AGNs are called 
``\textit{bare nucleus''} Seyferts or bare AGNs. The estimated mass of the central black hole 
of Ark 120 is M$_{BH}=1.50\pm0.19 \times 10^8$ M$_\odot$ \citep{Peterson2004} which was 
measured using the reverberation-mapping technique. From the spectroscopic monitoring data 
of Ark 120 during 1976 to 2013 using a 70 cm telescope, \cite{Denissyuk2015} estimated 
the mass of the central SMBH to be M$_{BH}=1.675\pm0.028 \times 10^8$ M$_\odot$. This source 
has a low Eddington ratio of $L_{bol}/L_E \sim  0.05$ \citep{Vasudevan2007} with a strong 
soft-excess \citep{Matt2014, Porquet2004, Porquet2019} and a significant broad Fe 
K$_\alpha$ line \citep{Vaughan2004, Nardini2011}. \cite{Nardini2011} analyzed Ark 120 
spectra, where, in the absence of absorber of complex morphology, soft-excess was explained by 
reflection from the centrally located hot and cold medium located at a distance. 
\cite{Marinucci2019} used the Monte-Carlo technique to investigate the favourable shape of the 
Compton cloud considering the future polarimetric missions such as {\it IXPE} \citep{Weisskopf2016}.

Although Ark 120 is a widely studied source, the evolution of the X-ray spectra over the 
last two decades is yet to be understood. However, a steepening of the X-ray spectrum was observed
during six-month monitoring in 2014 with Swift. The observed spectral variability was attributed to the 
possible existence of a large disc reprocessing region \citep{Gliozzi2017}. Again during 2017-18, 
a longer time delay was observed \citep{Lobban2018} between longer wavelength difference 
(i.e., optical and X-ray). They predicted that the accretion disc could exist in a longer scale 
as predicted by standard accretion disc theory. The soft-excess part of Ark 120 could be originated 
due to the Comptonization within the hot electron cloud of various shape \citep{Marinucci2019}, 
reflection from a cold medium \citep{Nardini2011} or the shock heating near the inner edge 
of the disc \citep{Fukumura2016}. We analyzed the long 
term X-ray archival data of Ark 120 which provides an ideal testbed to understand the soft-excess 
as well as its interaction with the harder (>2 keV) photons. Along with the observations, 
we perform Monte-Carlo simulations to find the effect of Comptonizaton within the energy 
range of soft-excess. We also study the X-ray variability of the 
source over a longer period and to calculate the approximate time-delays in X-ray bands. For the 
first time, we also find the flow and system parameters by fitting of the X-ray data.
The paper is structured in the following way: in Sec \ref{sec:obs}, we provide the details 
of the observational data and their reduction procedure. The results of the spectral and 
temporal analysis are presented in Sec \ref{sec:spec} and \ref{sec:time}. 
We discuss our findings in Sec \ref{sec:dis} and finally, draw our conclusions 
in Sec \ref{sec:con}.


\section{Observation and Data Reduction}
\label{sec:obs}
We use the publicly available archival data of {\it XMM-Newton}, {\it NuSTAR}, 
{\it Chandra}, and {\it Suzaku} using HEASARC\footnote{\url{http://heasarc.gsfc.nasa.gov/}}. 
We reprocessed all data using {\tt HEAsoft  v6.26.1} \citep{Arnaud1996}, which includes 
{\tt XSPEC v12.10.1f}.
	
\begin{table}
	\tiny{
	\caption{Observation Log}
	\label{tab:1}
	\begin{tabular}{lcccc}
		\hline
		ID&Date & Obs. ID & Instrument & Exposures\\
		  &(yyyy-mm-dd) &         &            &(ks)\\
		\hline
		XMM1 & 2003-08-24 & 0147190101  & {\it XMM-Newton}/EPIC-pn & 112.15 \\
		&&&&\\
		S1 & 2007-04-01 & 702014010   & {\it Suzaku}/XIS-HXD        & 100.86 \\
		&&&&\\
		XRT1 & 2008-07-24& 00037593001& {\it Swift}/XRT &  10.86\\
			&-2008-08-03& -00037593003& & \\
		&&&&\\		
		XMM2 & 2013-02-18 & 0693781501  & {\it XMM-Newton}/EPIC-pn & 130.46 \\
		&&&&\\
		N1   & 2013-02-18 & 60001044004 & {\it NuSTAR}/FPMA   & 65.46 \\
		&&&&\\
		XMM3 & 2014-03-22 & 0721600401  &{\it XMM-Newton}/EPIC-pn &  124.0  \\
		&&&&\\
		N2   & 2014-03-22 & 60001044002 & {\it NuSTAR}/FPMA   & 55.33 \\
		&&&&\\
		XRT2 & 2014-09-04& 00091909002&{\it Swift}/XRT   &   22.81  \\
		     &-2014-10-19& -00091909022& & \\
		&&&&\\
		XRT3 & 2014-10-22 & 00091909023&{\it Swift}/XRT  & 20.18  \\
		     &-2014-12-05 & -00091909044& & \\
		&&&&\\
		XRT4 & 2014-12-09 & 00091909045& {\it Swift}/XRT & 23.48 \\
		     &-2015-01-26 & -00091909068& & \\
		&&&&\\
		XRT5 & 2015-01-26 & 00091909069&{\it Swift}/XRT &  21.66  \\
		     &-2015-03-15 & -00091909090& & \\
		&&&&\\
		XRT6 & 2017-12-07 & 00010379001&{\it Swift}/XRT &  44.14  \\
		     &-2018-01-24 & -00010379048& & \\	
		\hline
	\end{tabular}
}
\end{table}

\subsection{XMM-Newton}
\label{sec:xmm}
Ark 120 has been observed by {\it XMM-Newton} \citep{Jansen2001} during three epochs
from 2003 to 2014. In 2003 and 2013, it has made $\sim$ 112 ks (XMM1) and $\sim$ 130
ks (XMM2) observations respectively. The XMM1 data is used by \citep{Vaughan2004} and
reported that the source Ark 120 is one of the cleanest Sy1 type AGN. In 2014, {\it
XMM-Newton} observed Ark 120 four times between March 18 and March 24. Out of these,
one (XMM3) was simultaneous with {\it NuSTAR} observation. The details
of the observation log are presented in Table~ \ref{tab:1}. It was observed that
the X-ray flux of this source was about a factor of two higher  in 2014 than the XMM2
observation \citep{Matt2014, Marinucci2019} made in 2013. A similar trend of flux 
variation was also reported in optical/UV \citep{Lobban2018} band. 

Due to the high brightness of the source, the European Photon Imaging Camera 
(EPIC-pn \citep{Struder2001}) operated in Small
Window (SW) mode to prevent any pile-up. The details of the {\it XMM-Newton}/EPIC-pn
observations of this source are listed in Table-\ref{tab:1}. We reprocessed the raw data to
level 1 data for EPIC-pn by Scientific Analysis System 
({\tt SAS v16.1.0}\footnote{\url{https://www.cosmos.esa.int/web/xmm-newton/sas-threads}}) with calibration
files dated February 2, 2018. We have used only the unflagged ({\tt FLAG == 0}) events
for excluding the edge of CCD and the edge of the bad pixel. Besides this, 
we also use {\tt PATTERN} $\leq 4$ for single and double pixel.
We exclude the photon flares by proper {\tt GTI} files to acquire the maximum signal to noise ratio.
After that, we use an annular area of 30\arcsec~ outer radii and 5\arcsec~ inner radii
centered at the source to extract the source event. For the background, we use a circle of
60\arcsec~ in the lower part of the window that contains no (or negligible) source photons.
The response files (\textit{arf} and \textit{rmf} files) for each EPIC-pn spectral data set
were produced with {\tt SAS} tasks {\tt ARFGEN} and {\tt RMFGEN}, respectively. The
{\tt GRPPHA} task is used with 100 counts per bin for 0.3 - 10.0 keV EPIC-pn spectra.

\subsection{Suzaku}
\label{sec:suzaku}
{\it Suzaku} observed  
Ark 120 on 2007 April 1 (Obs ID: 702014010) in HXD normal position with exposure
of $\sim$ 101 ks using X-ray imaging spectrometer \citep{Koyama2007} and $\sim$ 89 ks for
Hard X-ray Detector \citep{Takahashi2007}. The photons were collected in both $3\times3$ 
and $5\times5$ editing modes. From this observation, a presence of soft-excess emission
in soft X-ray was reported by \citep{Nardini2011}. Also, Fe K$\alpha$ emission line with
full-width at half maximum of $4700^{+2700}_{-1500}$ km s$^{-1}$ was previously 
reported by \citep{Nardini2016} by using {\it Suzaku} observation along with 
{\it XMM-Newton}, {\it Chandra}i, and {\it NuSTAR}.
	 
We use the standard data reduction technique for {\it Suzaku} data analysis illustrated in 
{\it Suzaku} Data Reduction Guide\footnote{\url{http://heasarc.gsfc.nasa.gov/docs/suzaku/analysis/abc/}} 
and followed the recommended screening criteria while extracting {\it Suzaku}/XIS spectrum and 
light-curves. The latest calibration files\footnote{\url{http://www.astro.isas.jaxa.jp/suzaku/caldb/}} 
available (2014-02-03) using {\tt FTOOLS 6.25} is used to reprocess the event files. 
The source spectra and lightcurves are extracted from a circular region of radius 
200\arcsec centered on the Ark 120 and the background region is selected on the same 
slit with a circular region 250\arcsec. Finally, we merge the two front illuminated 
detectors (XIS0 and XIS3) to produce the final spectra and lightcurves for Ark 120. We 
generated the response files through {\tt XISRESP} script. 
	
As {\it Suzaku} has a high energy X-ray detector (HXD), we use the HXD/PIN data for 
our analysis. We reprocessed the unfiltered event files using the standard tools. The 
output spectrum and lightcurves are extracted by using the {\tt hxdpinxbpi} and {\tt 
hxdpinxblc}, respectively. Further, we correct the spectrum to take into account both 
the non-X-ray and the cosmic X-ray backgrounds and the dead time correction.
	
\subsection{NuSTAR}
\label{sec:nustar}
{\it NuSTAR} \citep{Harrison2013} observed Ark 120 simultaneously with {\it XMM-Newton} 
with FPMA and FPMB on 2013 February 18 (N1) and 2014 March 22 (N2) for the exposure of 
$\sim$166 ks and $\sim$ 131 ks respectively. The details of the observation log are 
given in Table~ \ref{tab:1}. We consider both N1 and N2 observations for our analysis. 
\citep{Porquet2018, Porquet2019} used this data along with {\it XMM-Newton} and 
determined the spin $0.83^{+0.05}_{-0.03}$ and comment on the dimension of the corona and 
temperature by analyzing these X-ray data. 
	
The level 1 data is produced from the raw data by using the {\it NuSTAR} data analysis	
software ({\tt NuSTARDAS v1.8.0}). The cleaned event files are produced with standard 
{\tt NUPIPELINE} task and calibrated with the latest calibration files available in the 
NuSTAR calibration database
(CALDB)\footnote{\url{http://heasarc.gsfc.nasa.gov/FTP/caldb/data/nustar/fpm/}}. We 
chose 90\arcsec~ radii for source and 180\arcsec~ radii for the background region on the 
same detector to avoid contamination and detector edges. For the final 
background-subtracted lightcurves, we use 100s bin for both FPMA and FPMB. As both 
detectors are identical, here we present the results of FPMA only. The response files
(\textit{arf} and \textit{rmf} files) are generated by using the {\tt numkrmf} and 
{\tt numkarf} modules, respectively.  
	
\subsection{Swift data}
Swift X-ray telescope (XRT; \cite{Burrows2005}), working in the energy range of
0.2 to 10.0 keV, is an X-ray focusing telescope. XRT observed this source in both WT 
(windowed timing) and pc (photon count) modes depending on the brightness of the 
source. Ark 120 was observed over $\sim 130$ times from 2008-07-24 to 2018-01-24. 
In 2008, {\it Swift} observed three times, July 24, July 31 and August 3. 
We stack the spectra to produce a combined spectrum (XRT1). Then, it again observed 
on 2014-03-22, which has a simultaneous observation with {\it XMM} and {\it NuSTAR}. 
We consider the XMM3 observation over this particular XRT observation. {\it Swift} observed 
Ark 120 from 2014-09-06 to 2015-03-15 on a nearly daily basis. Further, we stack 
these observations into four observations (XRT2, XRT3, XRT4, XRT5) with each observations 
spanning around 50 days. In the last epoch, {\it Swift} observed Ark 120 from 2017-12-05 
to 24-01-2018 over $\sim$ 50 days. We stack the observations to produce the spectra 
of XRT6. The details of the observation log are stated in Table \ref{tab:1}. We use the 
online tool ``XRT product builder''\footnote{\url{http://swift.ac.uk/user_objects/}} 
\cite{Evans2009} to extract the spectrum and light curves. This product builder 
performs all necessary processing and calibration and produces the final spectra and 
lightcurves of Ark 120 in WT and PC mode.
    
\section{Spectral Analysis}
\label{sec:spec}
We use {\it XMM-Newton}, {\it Suzaku}, {\it NuSTAR}, and {\it Swift} data for the
spectral analysis and explore the spectral variation over $\sim$15 years 
(2003-2018) period using {\tt XSPEC v12.10.1f} \citep{Arnaud1996}. We explore 
the broad spectral properties with {\tt nthcomp} model \citep{Zdziarski1996}. 
Later, we apply Two Component Advective Flow (TCAF) model \citep{CT95} to 
extract the physical flow parameters such as the accretion rates and size 
of the Compton cloud.

Along with these models, we use a {\tt Gaussian} component for the Fe 
fluorescent emission line. While fitting the data, we use two absorption 
components, namely {\tt TBabs} and {\tt zTBabs} \citep{Wilms2000}. 
The component, {\tt TBabs} is used for the Galactic 
absorption, where hydrogen column density ($N_{H,gal}$) is fixed at
$9.78 \times 20^{20}$ cm$^{-2}$ \citep{Kalberla2005}. To calculate the error
for each parameter in spectral fitting with 90\% confidence level, we use 
`{\tt error}' command in {\tt XSPEC}.
	
We use following cosmological parameters in this work: $H_0$ = 70 km s$^{-1}$ Mpc
$^{-1}$, $\Lambda_0$ = 0.73, $\Omega_M$ = 0.27 \citep{Bennett2003}. With the assumed 
cosmological parameters, the luminosity distance of Ark 120 is 142 Mpc. 

\begin{table*}
	\caption{\label{tab:2} {\tt nthcomp} fitting result for the spectrum 
		above 3.0 keV. The optical depth $\tau$ is calculated from 
		equation-\ref{eq:1}.}
	\begin{center}
		\begin{tabular}{c c c c c c c c}
			\hline
ID   & MJD &$\Gamma^{nth} $  &$kT_e$  & Fe $K_\alpha$     & EW  & $\chi/dof$   &$\tau^*$\\
     &     &           &(keV)   &  (keV)            &(eV) &              &        \\
			\hline
XMM1        & $52875 $  &$1.90^{+0.01}_{-0.01} $ &$159.45^{+81.68}_{-81.69} $     &$6.40^{+0.016}_{-0.017}$ & $116^{+3}_{-4}$    & 312.33/300   &$0.733\pm0.003 $\\
			&&&&&&&\\                                                                                                                     
S1          & $54191 $  &$2.08^{+0.03}_{-0.03} $ &$124.65^{+35.54}_{-35.21} $     &$6.38^{+0.052}_{-0.052}$  & $710^{+10}_{-10}$ & 1117.31/1093 &$0.726\pm0.008 $\\
			&&&&&&&\\                                                                                                                     
XRT1        & $54676 $  &$1.76^{+0.02}_{-0.08} $ &$217.72^{+105.6}_{-112.5}$      & -                        &  -                & 75.68/74     &$0.671\pm0.030 $\\
			&&&&&&&\\                                                                                                                     
XMM2+N1     & $56341 $  &$1.75^{+0.01}_{-0.02} $ &$221.56^{+105.3}_{-107.5} $     & $6.42^{+0.061}_{-0.062}$ & $136^{+8}_{-9}$   & 644.55/641   &$0.670\pm0.074 $\\
			&&&&&&&\\                                                                                                                     
XMM3+N2     & $56738 $  &$1.87^{+0.01}_{-0.01} $ &$205.95^{+100.6}_{-99.87} $     & $6.37^{+0.052}_{-0.052} $& $227^{+12}_{-11}$ & 508.07/469   &$0.612\pm0.003 $\\
			&&&&&&&\\
XRT2        & $56926 $  &$1.60^{+0.01}_{-0.02} $ &$274.40^{+136.5}_{-130.8} $     & $- $                                   &  -  & 306.65/290   &$0.700\pm0.008 $\\			
			&&&&&&&\\                                                                                                                     
XRT3        & $56974 $  &$1.84^{+0.02}_{-0.02} $ &$215.72^{+105.5}_{-105.8} $     & $- $                                   &  -  & 319.98/320   &$0.610\pm0.006 $\\
			&&&&&&&\\                                                                                                                     
XRT4        & $57024 $  &$1.72^{+0.02}_{-0.03} $ &$225.57^{+109.7}_{-109.9} $     & $- $                                   &  -  & 269.17/280   &$0.688\pm0.011 $\\
			&&&&&&&\\                                                                                                                     
XRT5        & $57073 $  &$1.88^{+0.02}_{-0.02} $ &$201.58^{+99.78}_{-99.20} $     & $- $                                   &  -  & 246.53/261   &$0.616\pm0.006 $\\
			&&&&&&&\\                                                                                                                     
XRT6        & $58118 $  &$1.65^{+0.02}_{-0.02} $ &$246.87^{+120.9}_{-122.8} $     & $- $                                   &  -  & 327.78/318   &$0.708\pm0.008 $\\
			
			\hline
			
		\end{tabular}
	\end{center}
	
\end{table*}

\begin{table*}
\caption{\label{tab:SE} Soft-excess spectral indices are generated while keeping the
spectral slope of {\tt nthcomp} ($\Gamma^{nth}$) frozen. Intrinsic luminosities are calculated for both of the
components using {\tt clum} in the energy energy 0.5 to 10.0 kev.}
\begin{center}
\begin{tabular}{ c c c c c c c }
\hline
ID       & $\Gamma^{PC}$  & $Norm^{PC}$    &  $L^{PC}$               & $\Gamma^{SE}$          & $Norm^{SE}$            &  $L^{SE}$                \\
         & $=\Gamma^{nth}$& $(10^{-2})$    &                         &                        & $(10^{-2})$            &                          \\
\hline
&&&&&&\\
XMM1     & $1.90$ & $1.16^{+0.04}_{-0.05}$ & $44.18^{+0.06}_{-0.07}$ & $3.15^{+0.07}_{-0.06}$ & $0.58^{+0.02}_{-0.02}$ & $43.66^{+0.05}_{-0.04}$  \\
&&&&&&\\
S1       & $2.08$ & $18^{+20.3}_{-25.6}$   & $45.35^{+0.05}_{-0.05}$ & $2.52^{+0.02}_{-0.02}$ & $2105^{+10.6}_{-16.5}$ & $45.58^{+0.04}_{-0.04}$  \\
&&&&&&\\
XRT1     & $1.76$ & $0.66^{+0.03}_{-0.03}$ & $43.99^{+0.04}_{-0.04}$ & $4.11^{+0.22}_{-0.20}$ & $0.84^{+0.10}_{-0.10}$ & $43.87^{+0.02}_{-0.03}$  \\
&&&&&&\\
XMM2+N1  & $1.75$ & $0.57^{+0.03}_{-0.03}$ & $43.93^{+0.05}_{-0.05}$ & $3.03^{+0.03}_{-0.02}$ & $0.19^{+0.03}_{-0.05}$ & $43.16^{+0.03}_{-0.03}$  \\
&&&&&&\\
XMM3+N2  & $1.86$ & $1.21^{+0.01}_{-0.01}$ & $44.90^{+0.04}_{-0.04}$ & $4.23^{+0.02}_{-0.02}$ & $0.88^{+0.10}_{-0.10}$ & $43.19^{+0.04}_{-0.04}$  \\
&&&&&&\\
XRT2     & $1.60$ & $0.48^{+0.03}_{-0.03}$ & $44.92^{+0.04}_{-0.04}$ & $2.92^{+0.19}_{-0.20}$ & $0.57^{+0.03}_{-0.04}$ & $43.66^{+0.04}_{-0.04}$  \\
&&&&&&\\
XRT3     & $1.84$ & $0.79^{+0.03}_{-0.04}$ & $44.04^{+0.05}_{-0.05}$ & $3.27^{+0.27}_{-0.27}$ & $0.47^{+0.06}_{-0.06}$ & $43.57^{+0.05}_{-0.05}$  \\
&&&&&&\\
XRT4     & $1.72$ & $0.57^{+0.04}_{-0.05}$ & $43.94^{+0.04}_{-0.04}$ & $2.53^{+0.10}_{-0.12}$ & $0.39^{+0.05}_{-0.06}$ & $43.54^{+0.04}_{-0.04}$  \\
&&&&&&\\
XRT5     & $1.88$ & $0.76^{+0.03}_{-0.03}$ & $44.00^{+0.04}_{-0.04}$ & $3.17^{+0.34}_{-0.34}$ & $0.31^{+0.05}_{-0.05}$ & $43.37^{+0.05}_{-0.05}$  \\
&&&&&&\\
XRT6     & $1.65$ & $0.42^{+0.02}_{-0.03}$ & $43.84^{+0.03}_{-0.03}$ & $2.96^{+0.28}_{-0.29}$ & $0.23^{+0.02}_{-0.03}$ & $43.27^{+0.06}_{-0.05}$  \\
&&&&&&\\
\hline
\end{tabular}
\end{center}
\end{table*}

  \begin{table*}
  	\caption{\label{tab:3} The TCAF parameter space is defined in the file lmod.dat. }	
  	\vskip -0.2 cm
  	{\centerline{}}
  	\begin{center}
  		\begin{tabular}{c c c c c c c c}
  			\hline
\rm Model parameters & \rm Parameter units  &  \rm Default value  & \rm Min.  & Min.  &  Max.  &  Max.  & Increment \\
  			\hline
$\rm M_{BH}$&$\rm M_{Sun}$&$\rm 1.0\times10^8$&$2\times10^6$&$2\times10^6$&$5.5\times10^9$&$5.5\times10^9$&$ 10.0 $ \\
$\dot{m}_d $& $\rm Edd $ & $ \rm 0.001 $ & $ 0.0001 $ & $ 0.0001 $ & $ 1.0 $ & $ 2.0 $ & $ 0.0001 $ \\
$\dot{m}_h $& $\rm Edd $ & $ \rm 0.01 $ & $ 0.0001 $ & $ 0.0001 $ & $ 2.0 $ & $ 3.0 $ & $ 0.0001 $ \\
$\rm X_s $& $\rm r_g $ & $ \rm 100.0 $ & $ 10.0 $ & $ 10.0 $ & $ 1000.0 $ & $ 1000.0 $ & $ 2.0 $ \\
$\rm R $& $\rm  $ & $ \rm 1.5 $ & $ 1.1 $ & $ 1.1 $ & $ 6.8 $ & $ 6.8 $ & $ 0.1 $ \\
  			
  			\hline
  			
  		\end{tabular}
  	\end{center}
  \end{table*}

\subsection{{ Nthcomp}}
We have started the spectral fitting with {\tt nthcomp} model, and the model in 
{\tt XSPEC} reads as:

{\tt  TBabs*zTBabs*(nthcomp+zGaussian)} 

{\tt nthcomp} is a thermally Comptonized continuum model proposed by
\cite{Zdziarski1996} and later extended by \cite{Zycki1999}. We fit all 
X-ray spectrum above 3.0 keV by this baseline model. The model depends on the seed photon 
energy ($kT_{bb}$), which we consider at 3 eV for all spectrum. Although, \cite{Marinucci2019} 
considered $kT_{bb}$ at 15 eV. It is to be noted that, we vary $kT_{bb}$ from 1 eV to 50 eV, 
and failed to notice any deviation in the residuals of the fitted spectra. We consider these 
seed photons to be disc-blackbody type. For that, we have opted for the {\it inp-type} is 1 for 
all fit. For the spectral fitting, first, we consider the energy range 3.0 to 10.0 keV. 
The fitted asymptotic power-law photon index $\Gamma=1.90$, electron temperature $kT_e=159.45$ keV and 
an iron K$\alpha$ line at 6.40 keV with equivalent width (EW) of $116^{+3}_{-4}$ eV with reduced 
chi-square ($\chi^2/dof$)=1.04 for degrees of freedom (dof) = 300 is obtained. Next, we analyse the 
data from the 2007 {\it Suzaku} observation. We have combined the {\it Suzaku}/XIS observation with 
{\it Suzaku}/HXD and make a spectrum from 0.5 to 40.0 keV. But, we fit 3.0 to 40.0 keV 
spectrum using the baseline model. The fitted parameters are $\Gamma=2.08$, 
$kT_e=124.65$ keV and iron K$\alpha$ line at 6.38 keV with equivalent width (EW) of $710^{+10}_{-10}$ eV. We are
also in need of an additional {\tt powerlaw} and {\tt Gaussian} to take care 
of high energy (above 10.0 keV) spectrum and emission lines. We have obtained the reduced 
chi-square ($\chi^2/dof$)=1.02 for degrees of freedom (dof) = 1093 
for this fitting. We have fitted the combined spectrum of XMM2+N1 (MJD-56341) and XMM3+N2 
(MJD-56738) spectrum using this model for the energy range 3.0 to 79.0 keV with the model 
parameters such as $\Gamma=1.75$ \& $1.87$ and corresponding $kT_e=221.56$ \& 
$205.95$ respectively. We have applied a {\tt zGaussian} for 
a Fe K$\alpha$ line at $6.42^{+0.061}_{-0.062}$ \& 
$6.37^{+0.052}_{-0.052}$ keV with equivalent widths (EW) of $136^{+8}_{-9}$ \& $227^{+12}_{-11}$  
eV for these combined spectra and the ($\chi^2/dof$)=644.55/641 \& ($\chi^2/dof$)= 508.07/469 
respectively. Next, we analyse the data obtained from {\it Swift}/XRT observation for the 
energy range of 3.0 to 10.0 keV. Fe K$\alpha$ line is not detected for all the six XRT spectra. 
We have fitted the {\it Swift}/XRT spectra by removing {\tt Gaussian} component from the baseline 
model. The power-law index $\Gamma$ vary from 1.60 to 1.88 and the corresponding electron 
temperature $kT_e$ vary from 274.40 to 201.58 keV respectively. The {\tt nthcomp} model fitted 
spectral analysis result is presented in Table \ref{tab:2}. Furthermore, we calculate the 
optical depth for each observation using the formula:

\begin{equation}
\label{eq:1}
	\tau=\sqrt{\frac{9}{4}+\frac{3}{\theta_e(\Gamma+2)(\Gamma-1)}} - \frac{3}{2},
\end{equation}
by inverting the relation A1 is presented in \cite{Zdziarski1996}. Here, $\theta_e=\frac{kT_e}{m_ec^2}$ is 
the electron energy with respect to the rest mass energy. The value of optical depth 
$\tau$ for each observation is provided in Table \ref{tab:2}. The maximum error in 
optical depth is obtained from $\Delta \tau \sim (\frac{1}{2}\frac{\Delta \theta_e}{\theta_e} + 
\frac{\Delta \Gamma}{\Gamma})\times\tau$, where $\Delta \theta_e$ and $\Delta \Gamma$ are considered 
from the fitted errors presented in Table \ref{tab:2}. 

We address the soft-excess (< 3 keV) part by adding another {\tt powerlaw} component. 
We freeze the $\Gamma$ obtained earlier while fitting the primary continuum alone. The second power-law
 fits the soft-excess, and the results are presented in Table \ref{tab:SE}. It should be noted that 
the spectral index of soft-excess ($\Gamma^{SE}$) is higher than the spectral index of the primary continuum 
($\Gamma^{PC}$) for every observation. 

\begin{figure}
	\vskip 0.5cm
	\includegraphics[width=1.0\columnwidth]{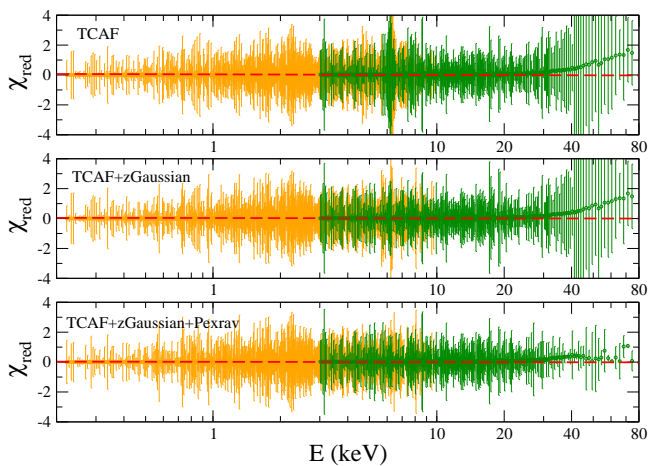}
	\caption{Variation of $\chi_{red}$ is shown for each model components on broadband spectra of 
	Ark 120 during 2014 epoch. Primarily, we have started with {\tt TCAF}, and then added {\tt zGaussian}
	and {\tt Pexrav} upon necessity.}
	\label{fig:chi_red}
\end{figure}
 
 \begin{figure*}
 	\begin{subfigure}[t]{0.33\textwidth}
 		\includegraphics[width=0.95\columnwidth]{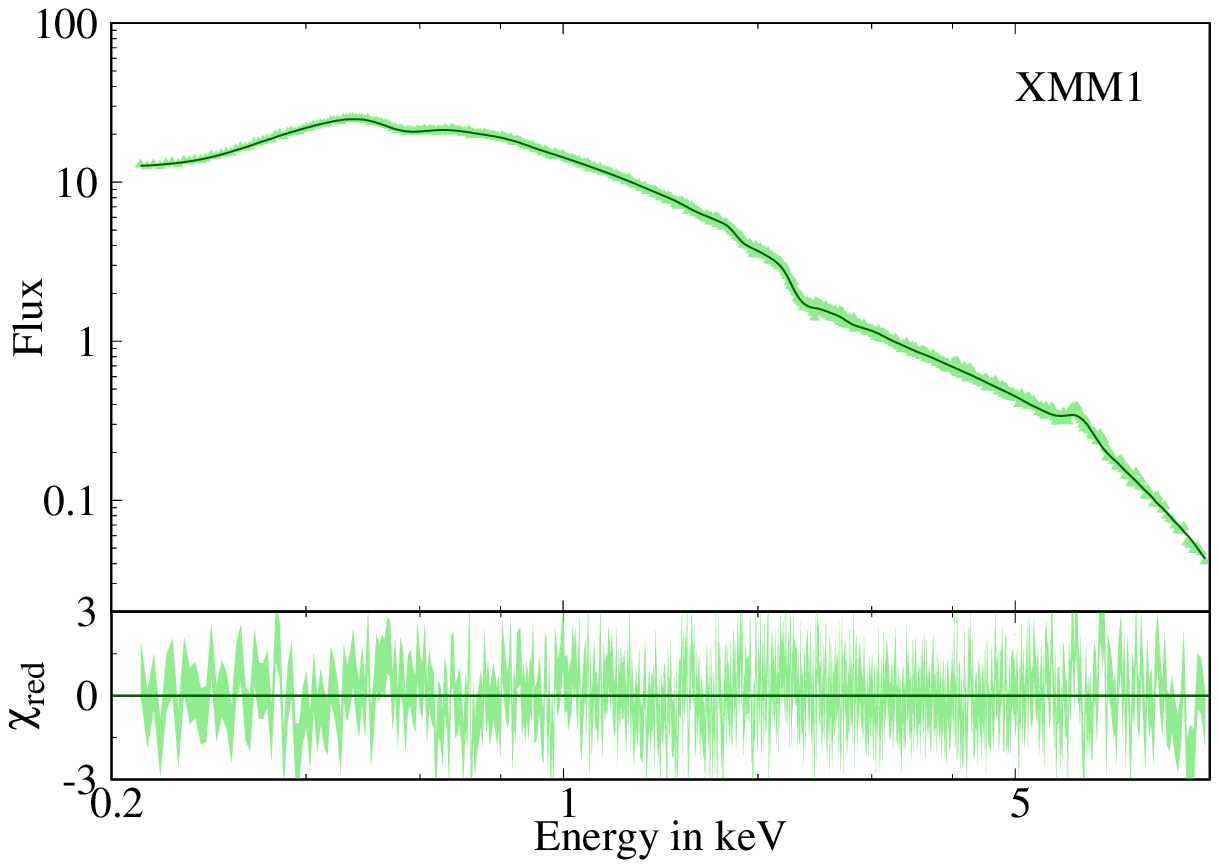}
 		\label{fig:6_1}
 	\end{subfigure}\hfill
 	\begin{subfigure}[t]{0.33\textwidth}
 		\includegraphics[width=0.95\columnwidth]{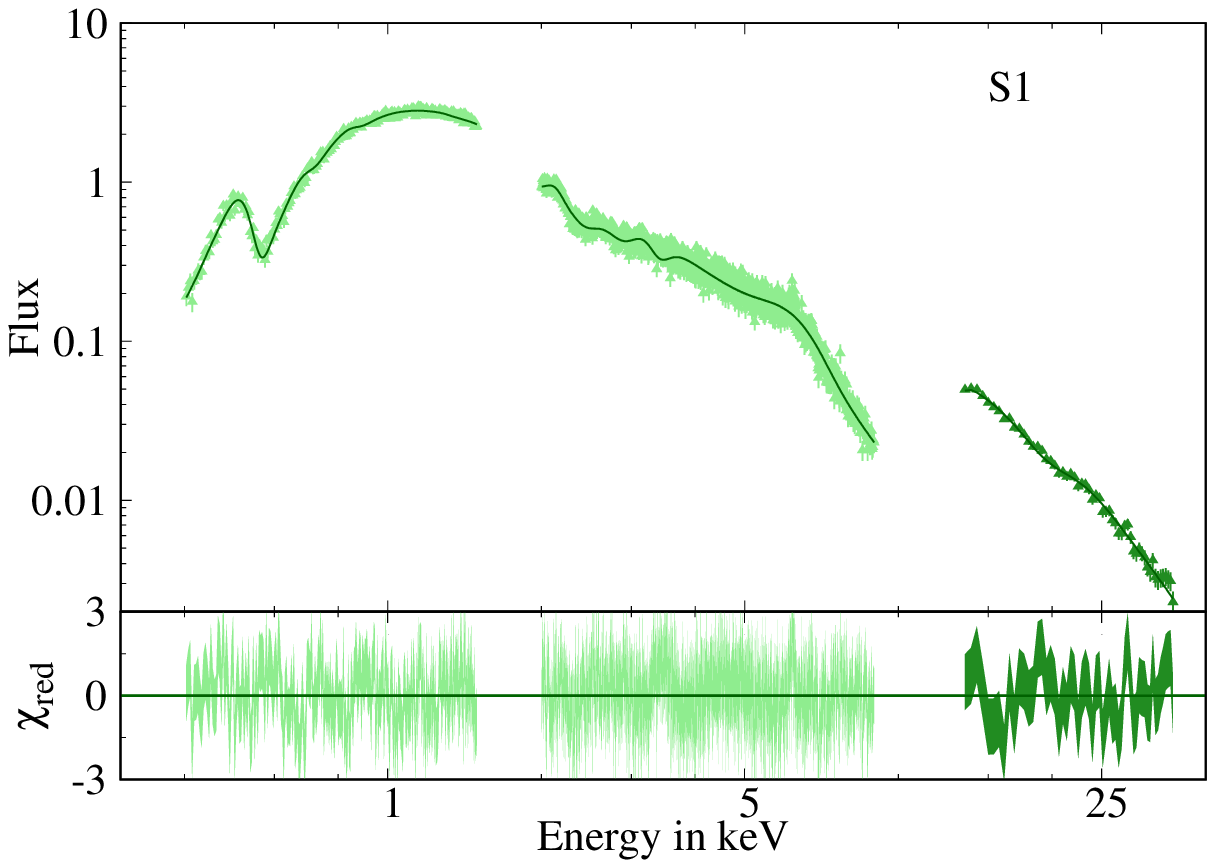}
 		\label{fig:6_2}
 	\end{subfigure}\hfill
 	\begin{subfigure}[t]{0.33\textwidth}
 		\includegraphics[width=0.95\columnwidth]{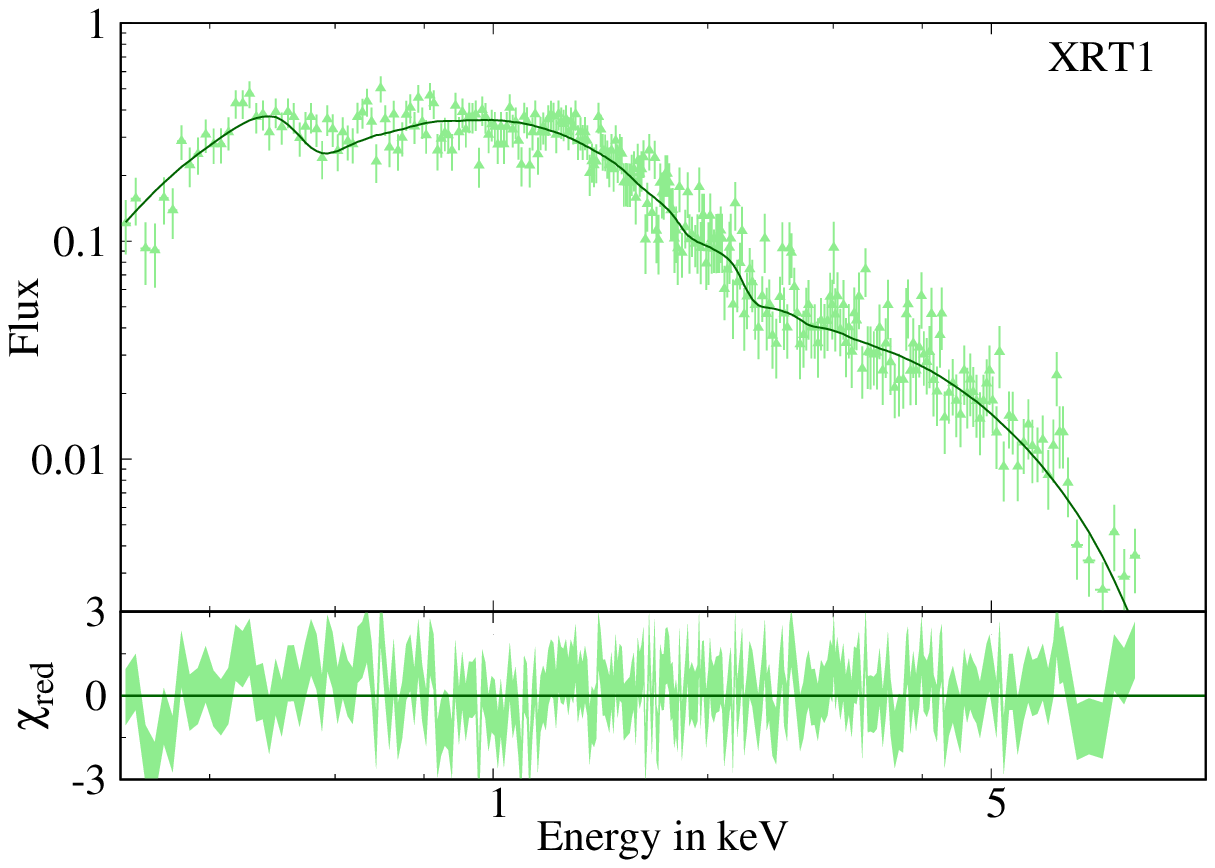}
 		\label{fig:6_3}
 	\end{subfigure}
 	\vskip 0.2cm
 	\begin{subfigure}[t]{0.33\textwidth}
 		\includegraphics[width=0.95\columnwidth]{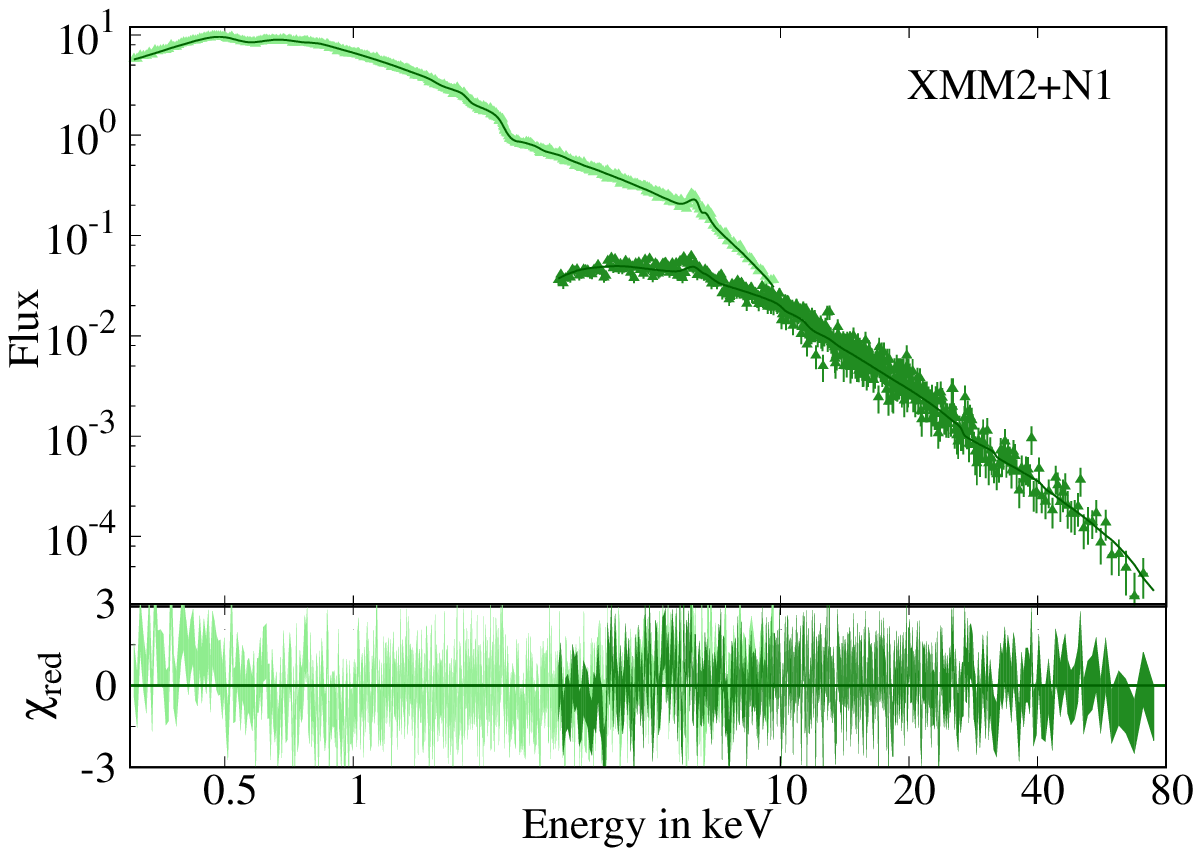}
 		\label{fig:6_4}
 	\end{subfigure}\hfill
 	\begin{subfigure}[t]{0.33\textwidth}
 		\includegraphics[width=0.95\columnwidth]{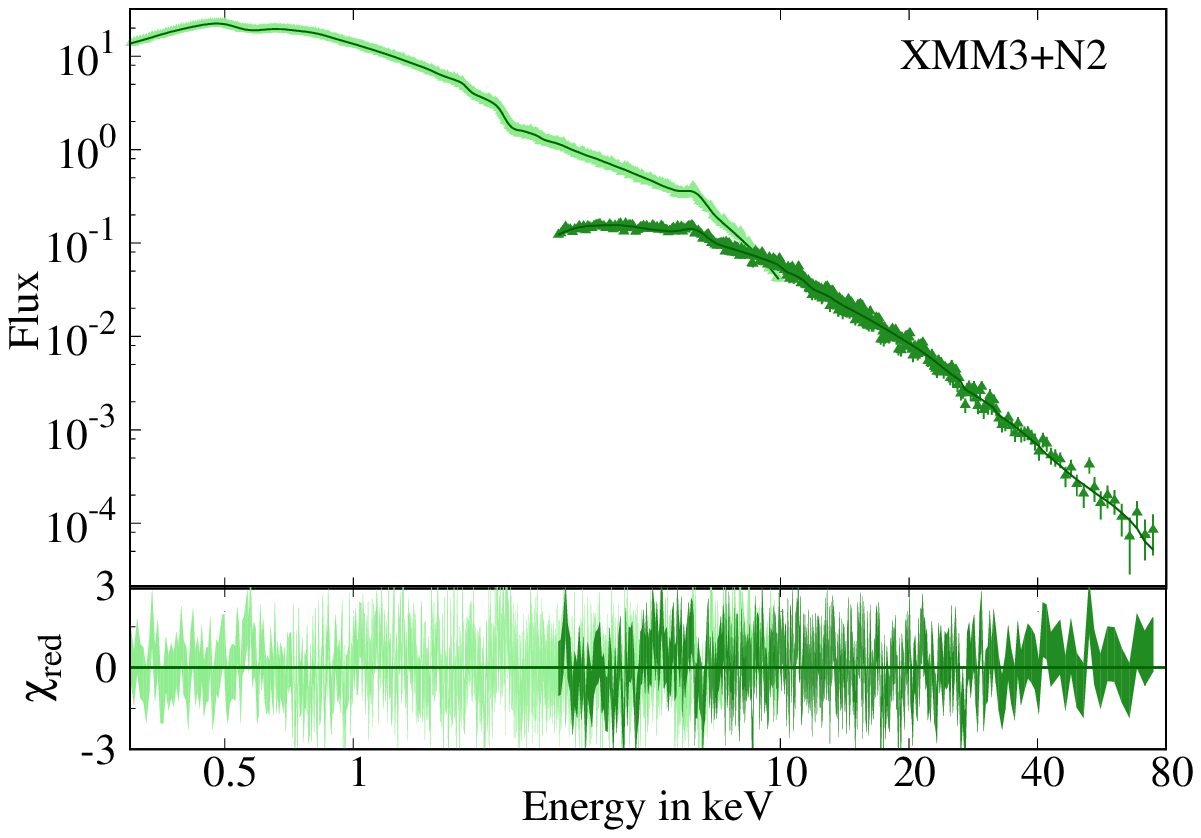}
 		\label{fig:6_5}
 	\end{subfigure}\hfill
 	\begin{subfigure}[t]{0.33\textwidth}
 		\includegraphics[width=0.95\columnwidth]{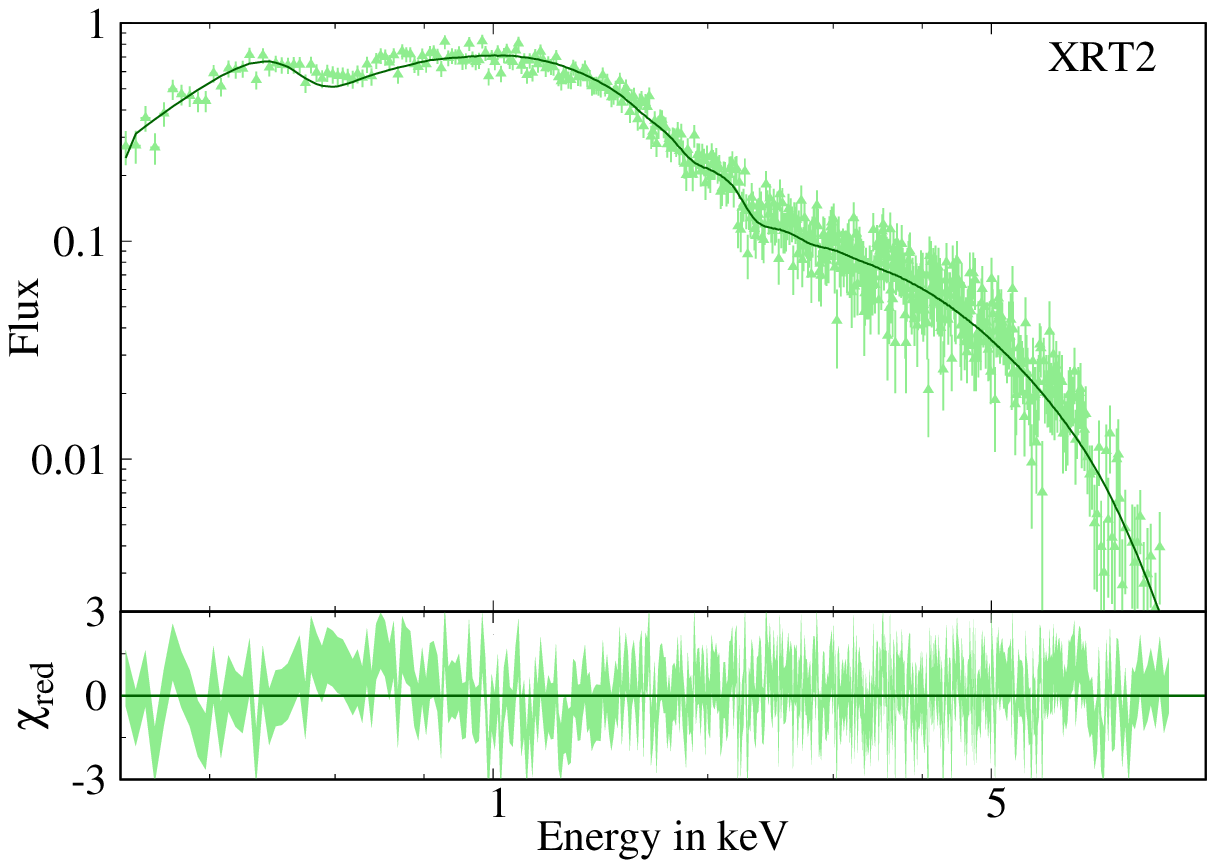}
 		\label{fig:6_6}
 	\end{subfigure}
 	\vskip 0.2cm
 	\begin{subfigure}[t]{0.33\textwidth}
 		\includegraphics[width=0.95\columnwidth]{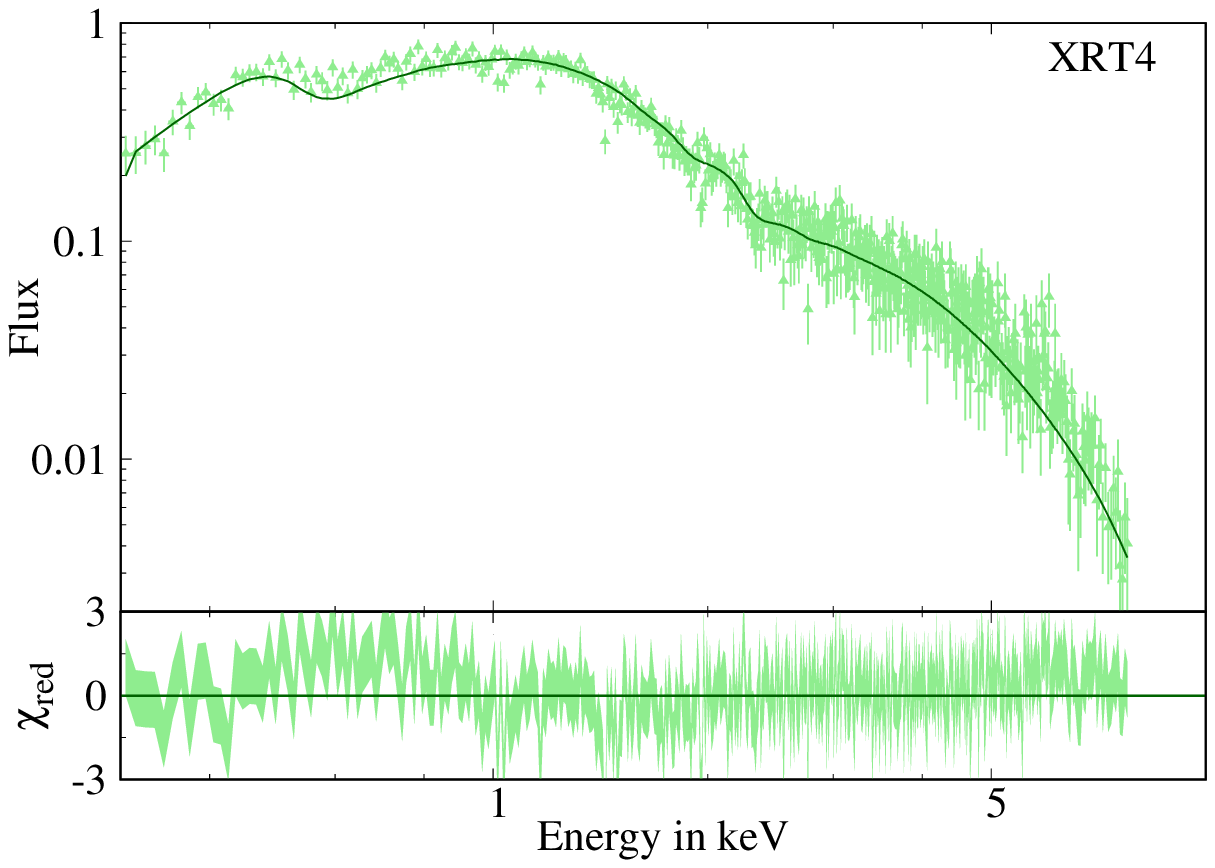}
 		\label{fig:6_7}
 	\end{subfigure}\hfill
 	\begin{subfigure}[t]{0.33\textwidth}
 		\includegraphics[width=0.95\columnwidth]{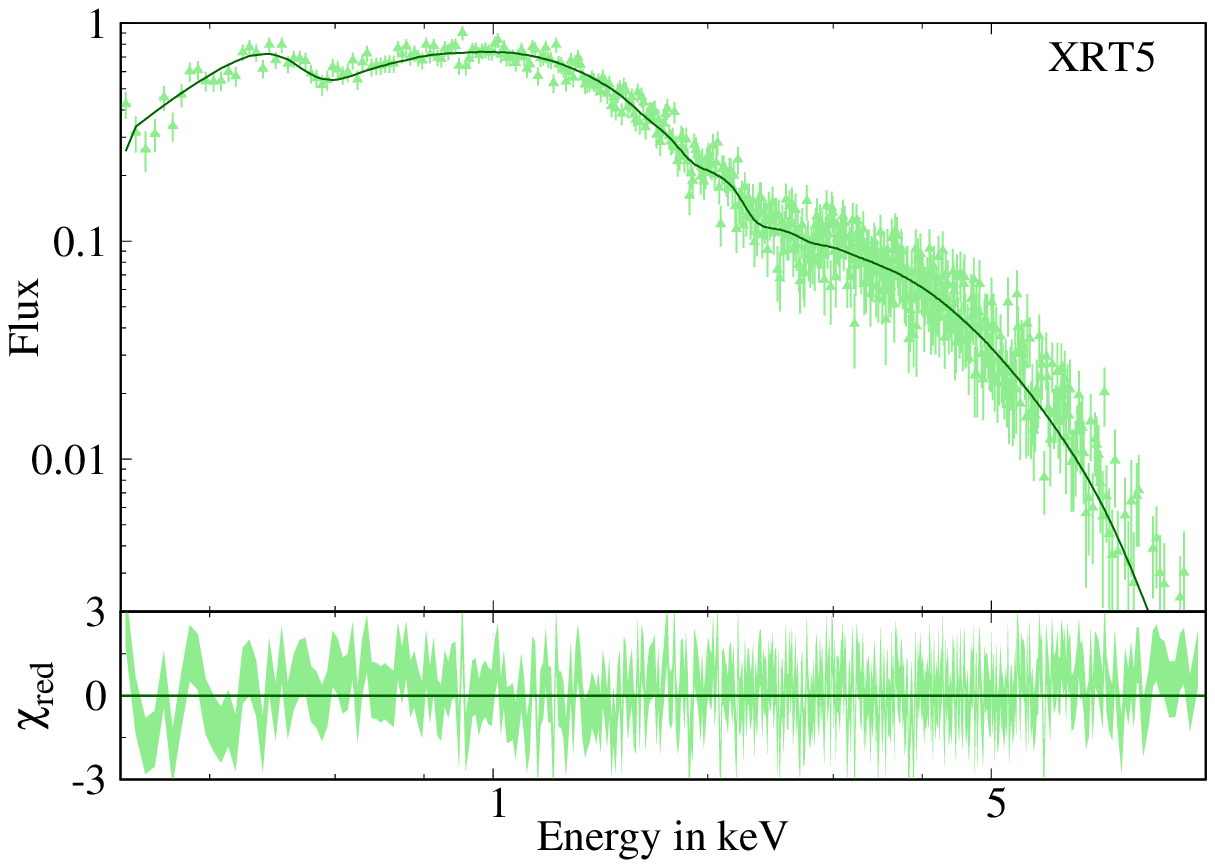}
 		\label{fig:6_8}
 	\end{subfigure}\hfill
 	\begin{subfigure}[t]{0.33\textwidth}
 		\includegraphics[width=0.95\columnwidth]{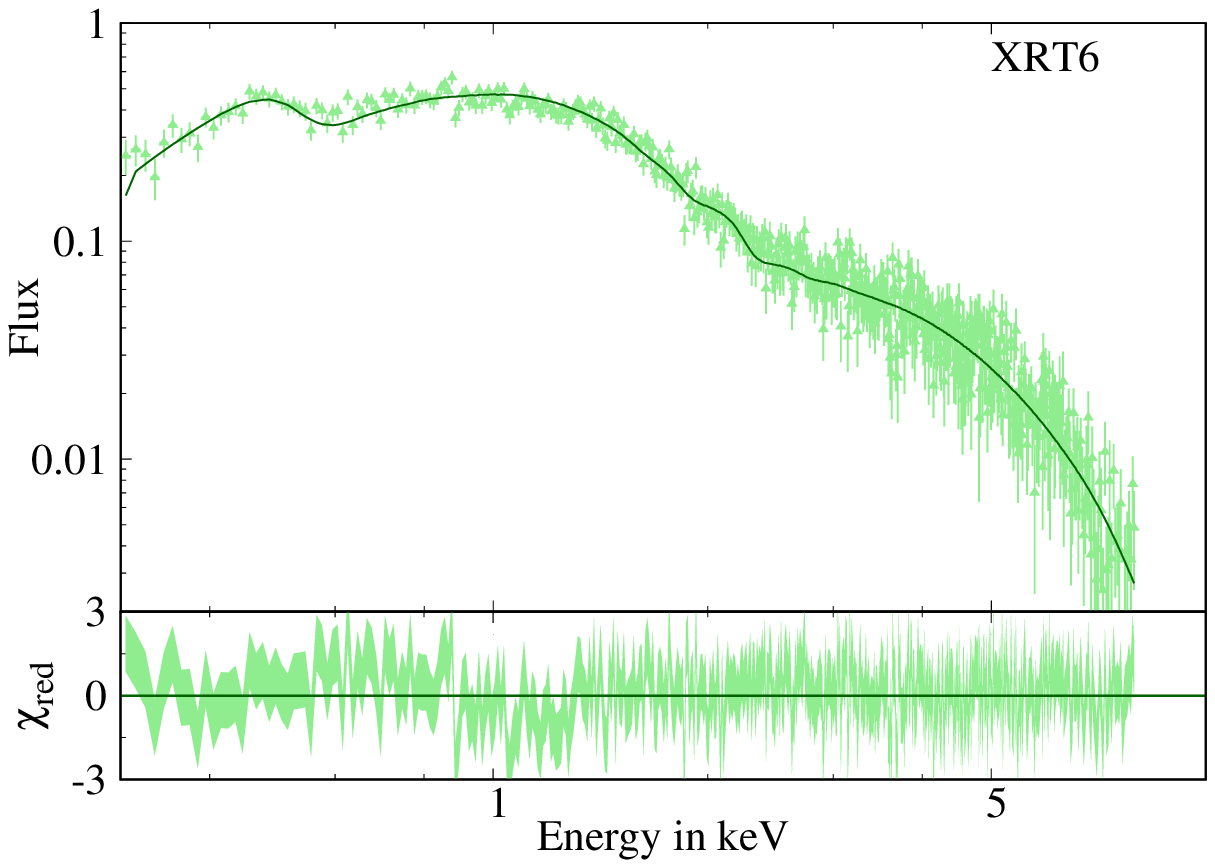}
 		\label{fig:6_9}
 	\end{subfigure}
 	
\caption{TCAF model fitted spectra of Ark 120 from the {\it XMM-Newton}, {\it Suzaku}, 
{\it NuSTAR} and {\it Swift} observations along with the residuals obtained from the 
spectral fitting.}
 	\label{fig:2}
 \end{figure*}


  \begin{table*}
  	\scriptsize
  	\caption{ {\tt TBabs*zTBabs*(TCAF+zGaussian)} model fitted Parameters 
  		in 0.2-79.0~keV energy band for Ark 120. The {\tt TBabs} is fixed 
  		at $N_{H_{gal}}$= $9.78\times10^{20}$ cm$^{-2}$. The second column 
  		shows the variation of {\tt zTBabs} for z = 0.033.}	
  	\vskip -0.2 cm
  	{\centerline{}}
  	\begin{center}
  		\begin{tabular}{c c c c c c c c c c c c c}
  			\hline
ID& MJD&$N_H$&$M_{BH}$ & $\dot{m}_d$ & $\dot{m}_h$ & $X_s$ & $R$ & $N_{TCAF}$ & $\Gamma_{pexrav}$ & $R_{ref}$ & $N_{pexrav}$ & $\chi^2/dof$\\
&    &$(10^{20}cm^{-2})$&$\rm (10^{8}M_\odot)$&$(\dot{m}_{Edd})$&$(\dot{m}_{Edd})$&$(r_g)$&  & $(10^{-3})$ &   & & ($10^{-3}$)&  \\
  			\hline
  			&&&&&&&&&&&& \\
XMM1&$\rm52875  $& $1.0^{+0.2}_{-0.2} $ &$ 1.50^{+0.03}_{-0.03} $& $\rm0.063^{+0.002}_{-0.002}  $ & $ \rm 0.112^{+0.001}_{-0.001} $ & $ 20.36^{+4.46}_{-4.55} $ & $ 1.95^{+0.34}_{-0.33} $  & $ 0.16^{+0.05}_{-0.05} $ & $ 0.14^{+0.10}_{-0.08}$ & $ 1.96^{+0.05}_{-0.09}$ & $ 0.14^{+0.04}_{-0.05}$  & $1026.20/842$\\
  			$ $& $ $&$ $&$ $&$ $&$ $&$ $&$ $&$ $&$ $& & & \\
S1&$\rm54191  $& $1.4^{+0.1}_{-0.2}  $ & $ 1.49^{+0.04}_{-0.04} $& $\rm0.126^{+0.002}_{-0.001}  $ & $ \rm 0.191^{+0.001}_{-0.001} $ & $ 21.44^{+4.96}_{-4.85} $ & $ 1.66^{+0.54}_{-0.57} $  & $ 233.6^{+54.26}_{-58.65} $ & $ 1.46^{+0.15}_{-0.41}$ & $ 0.642^{+0.71}_{-0.51}$  & $ 5.09^{+1.05}_{-1.03}$& $1869.89/1673$\\
  			$ $& $ $&$ $&$ $&$ $&$ $&$ $&$ $&$ $&$ $& & $ $ & \\
XRT1&$\rm54675  $& $1.6^{+0.6}_{-0.4}  $ & $ 1.49^{+0.19}_{-0.15} $& $\rm0.064^{+0.005}_{-0.005}  $ & $ \rm 0.110^{+0.004}_{-0.003} $ & $ 30.08^{+5.36}_{-6.24} $ & $ 2.80^{+0.54}_{-0.56} $  & $ 0.54^{+0.01}_{-0.02} $ & $ -$  & $ -$ & $ -$ & $307.78/283$\\
  			$ $& $ $& $ $ &$ $&$ $&$ $&$ $&$ $&$ $&$ $&$ $& & \\
XMM2+N1&$\rm54341  $& $1.9^{+0.3}_{-0.2}  $ & $ 1.50^{+0.08}_{-0.07} $& $\rm0.068^{+0.006}_{-0.005}  $ & $ \rm 0.111^{+0.004}_{-0.004} $ & $ 52.83^{+8.65}_{-8.56} $ & $ 2.83^{+0.55}_{-0.52} $  & $ 0.57^{+0.02}_{-0.02} $ & $ 0.90^{+0.09}_{-0.10}$ & $ 0.254^{+0.47}_{-0.06}$  & $0.25^{+0.06}_{-0.05}$& $1230.15/1112$\\
  			$ $& $ $& $ $ &$ $&$ $&$ $&$ $&$ $&$ $&$ $&$ $& & \\
XMM3+N2& $\rm54738  $& $2.3^{+0.3}_{-0.3}  $ & $ 1.51^{+0.09}_{-0.10} $& $\rm0.103^{+0.006}_{-0.005}  $ & $ \rm 0.126^{+0.004}_{-0.004} $ & $ 28.24^{+5.04}_{-5.25} $ & $ 2.43^{+0.55}_{-0.58} $  & $ 0.31^{+0.01}_{-0.01} $ & $ 1.66^{+0.19}_{-0.19}$ & $ 0.96^{+0.05}_{-0.56}$ & $0.12^{+0.06}_{-0.06}$ & $1578.92/1359$\\
  			$ $& $ $& $ $ &$ $&$ $&$ $&$ $&$ $&$ $&$ $&$ $& &\\
XRT2& $\rm56926  $& $2.5^{+0.1}_{-0.1}  $ & $ 1.49^{+0.18}_{-0.20} $& $\rm0.068^{+0.005}_{-0.006}  $ & $ \rm 0.110^{+0.003}_{-0.003} $ & $ 53.56^{+8.27}_{-8.87} $ & $ 2.73^{+0.51}_{-0.47} $  & $ 0.36^{+0.01}_{-0.02} $ & $ -$ & $ -$ & $ -$ & $579.89/555$\\
  			$ $& $ $& $ $ &$ $&$ $&$ $&$ $&$ $&$ $&$ $&$ $&& \\
XRT3&$\rm56974  $& $1.9^{+0.5}_{-0.5}  $ & $ 1.51^{+0.19}_{-0.20} $& $\rm0.068^{+0.006}_{-0.006}  $ & $ \rm 0.110^{+0.004}_{-0.005} $ & $ 55.16^{+8.57}_{-8.80} $ & $ 2.74^{+0.45}_{-0.41} $  & $ 0.25^{+0.01}_{-0.01} $ & $ -$ & $ -$ & $ -$ & $630.08/594$\\
  			$ $& $ $& $ $ &$ $&$ $&$ $&$ $&$ $&$ $&$ $&$ $& &\\
XRT4&$\rm57024  $& $1.1^{+0.2}_{-0.3}  $ & $ 1.50^{+0.15}_{-0.18} $& $\rm0.061^{+0.006}_{-0.006}  $ & $ \rm 0.110^{+0.005}_{-0.003} $ & $ 56.86^{+10.97}_{-10.89} $ & $ 2.69^{+0.48}_{-0.47} $  & $ 0.11^{+0.01}_{-0.01} $ & $ -$ & $ -$ & $ -$ & $702.39/548$\\
  			$ $& $ $& $  $ &$ $&$ $&$ $&$ $&$ $&$ $&$ $&$ $& &\\
XRT5& $\rm57073  $& $1.4^{+0.4}_{-0.3}  $ & $ 1.49^{+0.16}_{-0.15} $& $\rm0.069^{+0.006}_{-0.007}  $ & $ \rm 0.110^{+0.005}_{-0.005} $ & $ 57.87^{+12.99}_{-12.08} $ & $ 2.77^{+0.52}_{-0.54} $  & $ 0.28^{+0.02}_{-0.01} $ & $ -$ & $ -$ & $ -$ & $551.49/531$\\
  			$ $& $ $& $ $ &$ $&$ $&$ $&$ $&$ $&$ $&$ $&$ $& &\\
XRT6& $\rm58118  $& $2.0^{+0.4}_{-0.4}  $ & $ 1.51^{+0.16}_{-0.15} $& $\rm0.081^{+0.006}_{-0.005}  $ & $ \rm 0.140^{+0.006}_{-0.007} $ & $42.95^{+8.98}_{-8.20} $ & $ 2.69^{+0.59}_{-0.61} $  & $ 0.22^{+0.01}_{-0.01} $ & $ -$ & $ -$ & $ -$  & $612.9/589$\\
  			$ $& $ $& $ $ &$ $&$ $&$ $&$ $&$ $&$ $&$ $&$ $& &\\
  			\hline
  		\end{tabular}
  	\end{center}
  	\label{tab:4}
  \end{table*}

\begin{figure*}
	\vskip 1.5cm
	\includegraphics[width=2.0\columnwidth]{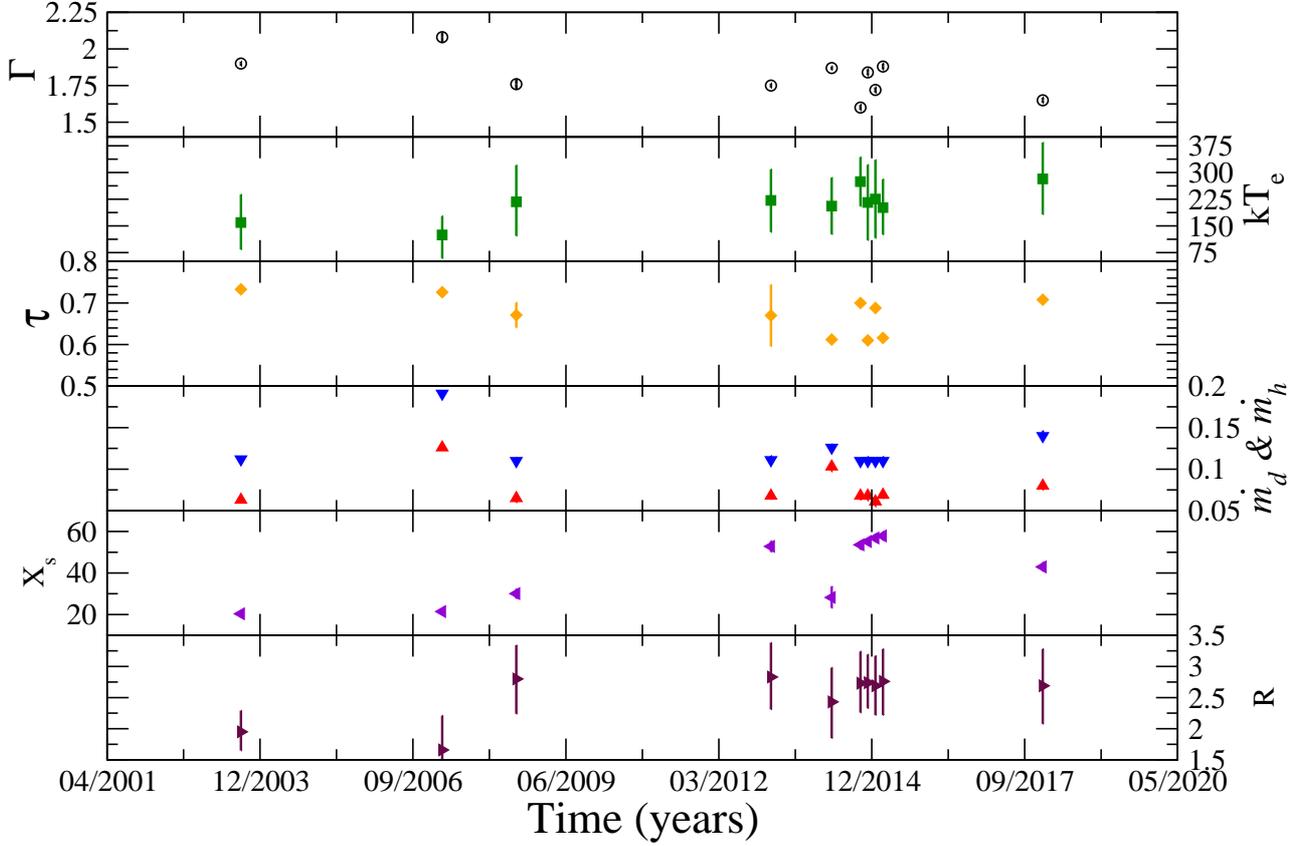}
	\caption{Variation of different model parameters with time are presented.} 
	\label{fig:evolve}
\end{figure*}

\subsection{TCAF}
\label{sec:TCAF}
From the {\tt nthcomp} model fitting, we have extracted several valuable information
on the spectral hardness and electron temperature of the emitting system
in a time duration of $\sim$ 15 years. We have also calculated the optical depths
from these parameters, which are shown in Table \ref{tab:2}. However, the fundamental 
properties, such as the central black hole mass, accretion rates, the size of the Compton 
cloud radius could provide a deeper physical understanding of the system. To 
estimate these quantities, we use the Two-Component Advective Flow (TCAF) model 
\citep{CT95} for our spectral analysis. For the spectral fitting, the model 
in {\tt XSPEC} reads as: 

{\tt  TBabs*zTBabs*(TCAF+zGaussian)} 

{\tt TCAF} is based on one black hole parameter and four flow parameters: (i) black hole mass in units of 
the solar mass ($M_\odot$); (ii) Keplerian disc accretion rate 
($\dot{m}_d$) in the unit of the Eddington rate ($\dot{M}_{EDD}$);
(iii) Sub-Keplerian halo accretion rate ($\dot{m}_h$) in units of Eddington 
rate ($\dot{M}_{EDD}$); (iv) shock compression ratio (R) and
(v) shock location ($X_s$) in units of the Schwarzschild radius ($r_g=2GM/c^2$).
The upper and lower limits of all the parameters are put in a data
file called {\it lmodel.dat} provided in Table-\ref{tab:3} as an input to run the source 
code using {\tt initpackage} and {\tt lmod} task in {\tt XSPEC}. For the 
final spectral fitting of a specified observation, we run the source code for a 
vast number of times and select the best spectrum from many spectra using 
minimization of $\chi$ method. First, we have started fitting by the baseline 
model described as above. Some spectra, like XMM1, S1, XMM2+N1, XMM3+N2 have 
high reduced $\chi^2$ ($\chi^2_{red}>2$) value. We noticed that the model 
has deviated from the actual data at the high energy end. To compensate for that, 
we have added a {\tt powerlaw/pexrav} with the baseline model. Thus the model became: 

{\tt  TBabs*zTBabs*(TCAF+powerlaw/pexrav+zGaussian)}.

We have fitted the spectra with this model and found $\chi^2_{red}\approx1$. 
Further, to investigate the source of this power-law (whether 
it is from reflection or not), we have replaced the {\tt powerlaw} component by 
{\tt pexrav} \citep{MZ1995}. The {\tt pexrav} model has a power-law
continuum with a reflected component from an infinite neutral slab. We
have estimated the relative reflection coefficient ($R_{ref}$) with photon 
index ($\Gamma_{pexrav}$) and cosine of inclination angle $\cos \theta$ 
from the model fitting. We find $\theta$ to vary from $40\degree$ to 
$72\degree$. We fix abundances for heavy elements, such as iron at the
Solar value (i.e., 1). For the photon index ($\Gamma_{pexrav}$), first, 
we freeze its value to the value of $\Gamma$ obtained from {\tt nthcomp}. 
For this, we have found $\chi^2_{red}>2$. Thereafter, we {\tt thaw} this parameter 
and fit it again which have resulted $\chi^2_{red}\approx1$ with new value of 
$\Gamma_{pexrav}$.

We first fit the {\it XMM-Newton} observation (XMM1) during 2003 (MJD-52875) in the 
energy range of 0.2 to 10.0 keV with {\tt TBabs*zTBabs*(TCAF+zGaussian)} model.
However, we found a high $\chi^2_{red}$. The model has deviated after 9.2 keV from
the actual data. As mentioned above, we then add a {\tt powerlaw} with 
the baseline model, and then the {\tt powerlaw} is replaced by {\tt pexrav}. 
The fitted parameters are, $M_{BH}=1.5\times10^8 M_\odot$, $\dot{m}_d=0.063$, 
$\dot{m}_h=0.112$, $X_s=20.36$, $R=1.95$ with $\Gamma_{pexrav}=0.14$, $R_{ref}=1.96$, 
and $E_{fold}=16.08$ keV and the corresponding $\chi^2=1026.20$ with degrees of 
freedom (dof)= 842. The Fe line is found at $6.4$ keV with an equivalent width of 
$116$ eV.

Next, we consider the {\it Suzaku} observation (S1) of 2007 (MJD-54191). We 
combine the {\it Suzaku}/XIS and {\it Suzaku}/HXD spectra and make a
broadband spectrum in the energy range of 0.5 to 40 keV. We follow the similar
steps as described in XMM1 fitting and the fitted parameters are 
$M_{BH}=1.49\times10^8 M_\odot$, $\dot{m}_d=0.126$, $\dot{m}_h=0.191$, $X_s=21.44$, $R=1.66$
with $\Gamma_{pexrav}=1.46$, $R_{ref}=0.642$, and the corresponding 
$\chi^2_{red}/dof=1869.89/1673$. The position of 
Fe line is $6.38$ keV with an equivalent width of $710$ eV. It is to be noted that, 
within $6-7$ keV range, \cite{Nardini2011} reported the possibility of 
three lines for XMM1 and two lines for S1 observation respectively. 

Following a similar procedure, we fit the broadband spectra of Ark 120 for
the observations during 2013 XMM2+N1 (MJD-56341) and 2014 XMM3+N2 (MJD-56738). 
For these, we have obtained $M_{BH}=1.50$ \& $1.51\times10^8 M_\odot$, 
$\dot{m}_d=0.068$ \& $0.103$, $\dot{m}_h=0.111$ \& $0.126$, $X_s=52.83$ \& $28.24$, 
$R=2.83$ \& $2.43$ with $\Gamma_{pexrav}=0.96$ \& $1.66$ respectively. The details 
of data fitting are given in Table-\ref{tab:4}.
 
We fit all the six {\it Swift}/XRT spectra using the baseline model. Here, 
we do not find any Fe line in all these spectra. From the fitting, it 
is noticed that the mass of the central black hole $M_{BH}$ 
=$1.5\times10^8 M_\odot$, the disc $\dot{m}_d\sim 0.065$ and halo 
accretion rates $\dot{m}_h$$\sim0.110$ are more or less constant 
except XRT6 observation. Here, we find $\dot{m}_d=0.081$ \& $\dot{m}_h=0.14$ 
and the corresponding shock location has moved inward from 57.87 to 
42.95 $r_g$. Therefore, the shock location ($X_s$) has varied in 
between $30.0$ to $57.87~r_g$, and the corresponding variation of 
the compression ratio (R) is in between $2.6$ to $2.8$ within September 2014 to 
January 2018. Here, we do not require any additional {\tt powerlaw}
to fit the high energy 
spectra. The details of the parameter variations are presented in Table-\ref{tab:4}. In Figure 
\ref{fig:2}, we plot the model fitted spectrum with the variation of $\chi$. Detailed 
discussions on spectral properties are demonstrated in Sec \ref{sec:pc}.
  
\begin{table*}
\centering
\caption{Variability statistics in various energy ranges are shown in this Table. We have opted 
for 100s time bins for variability analysis. In some cases, the average error of observational 
data exceeds the limit of $1\sigma$, resulting negative excess variance. In such cases, we 
have imaginary $F_{var}$, which are not shown in the table.}
\begin{tabular}{lccccccc}
\hline
ID      & Energy band     &    $N$   &$x_{max}$ &$x_{min}$&$\frac{x_{max}}{x_{min}}$   &$\sigma^2_{NXS}$      &$F_{var}   $\\
&&&&&&\\
	&   keV           &          & Count/s  & Count/s &      &$(10^{-2})     $      & $(10^{-2})$\\
\hline
XMM1    &0.2-2.0          &    1117   &21.95     &19.58    &1.12 & $0.57\pm0.003$       & $1.6\pm0.14$ \\
XMM2    &0.2-2.0          &    1294   &10.24     &8.40     &1.21 & $2.9\pm0.015 $       & $5.6\pm0.40$ \\
XMM3    &0.2-2.0          &    1309   &21.37     &17.11    &1.25 & $8.22\pm0.011$       & $6.4\pm0.41$ \\
\hline
XMM1    &3-10.0           &    1117   &1.95      &1.53     &1.79 & $0.18\pm0.023 $       & $3.2\pm0.04$ \\
XMM2    &3-10.0           &    1294   &1.22      &0.94     &1.30 & $0.137\pm0.035$       & $3.5\pm0.55$ \\
XMM3    &3-10.0           &    1309   &3.64      &2.92     &1.24 & $0.56\pm0.0011$       & $4.1\pm0.34$ \\
\hline
N1      &10.0-78.0        &    722    &0.711     &0.105    &6.748 & $0.093\pm0.007$      & $7.5\pm1.6$ \\
N2      &10.0-78.0        &    667    &1.712     &0.239    &7.143 & $0.147\pm0.024$      & $3.5\pm3.6$ \\
\hline
XMM1    &0.5-10.0         &    1117  &20.46     &15.26    &1.341 & $10.11\pm0.06$       & $2.36\pm0.14$ \\
S1      &0.5-10.0         &     586  &9.80      &5.58     &1.76  & $5.64\pm0.31$        & $8.62\pm0.31$ \\
XRT1    &0.5-10.0         &     8    &3.22      &1.67     &3.01  & -$22.3\pm5.6$        & $-$ \\
XMM2    &0.5-10.0         &    1294  &10.36     &6.73     &1.54  &  $2.93\pm0.02$       & $5.85\pm0.19$ \\
XMM3    &0.5-10.0         &    1309  &20.13     &13.67    &1.47  &  $5.80\pm0.01$       & $5.79\pm0.15$ \\
XRT2    &0.5-10.0         &      50  &2.13      &0.70     &3.02  &  $5.50\pm0.44$       & $20.70\pm2.3$ \\
XRT3    &0.5-10.0         &      43  &2.79      &1.52     &2.79  &  $3.61\pm0.42$       & $15.31\pm2.1$ \\
XRT4    &0.5-10.0         &      43  &1.90      &0.88     &2.16  &  $4.50\pm0.46$       & $18.01\pm2.3$ \\
XRT5    &0.5-10.0         &      42  &1.77      &0.81     &2.18  &  $2.72\pm0.25$       & $14.03\pm1.8$ \\
XRT6    &0.5-10.0         &      72  &1.63      &0.52     &3.09  &  $6.02\pm0.49$       & $23.40\pm2.2$ \\
\hline
\end{tabular}
\label{tab:fvar}
\end{table*}

\begin{figure*}
	\includegraphics[width=2.0\columnwidth]{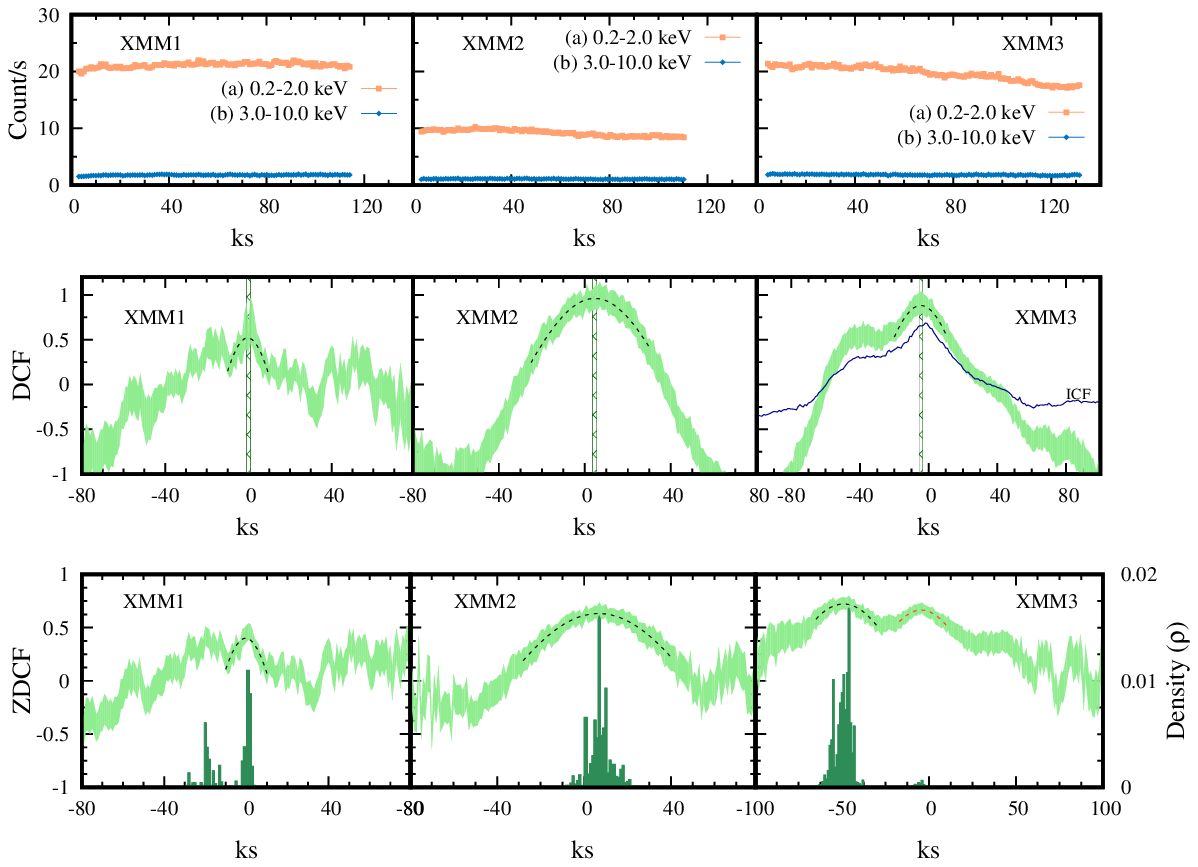}
	\caption{{\it Top panel:} The light-curves of the energy ranges of $0.2$ to $2.0$ keV and 
	$3.0$ to $10.0$ keV observed 
	by {\it XMM-Newton} are plotted for three epochs. The high energy count always remained a
	fraction of low energy counterpart. In 2013, the low-energy count dropped to nearly 50\% 
	as compared to 2003. Again in 2014, the $0.2-2$ keV count doubled from its value observed in 2013. 
	{\it Middle panel:} Corresponding discrete cross-correlations between light-curves of $0.2-2$ keV 
	and $3-10$ keV are plotted. All three epochs exhibited different patterns where {\it zero}, 
	{\it positive}, and {\it negative} delays are observed in 2003, 2013, and 2014 respectively. 
	We have also presented the {\tt ICF} (solid-blue line) for XMM3 observation.
	{\it Lower panel:} $\zeta$-discrete cross-correlations (light-green) are plotted for 
	light-curves of $0.2-2$ keV and $3-10$ keV. While 2003 and 2013 patterns remain similar to what
	have been observed from {\tt DCF}, the pattern obtained from 2014 data develops twin peak. 
	The likelihoods (dark-green), simulated using 12000 points, are plotted along with the {\tt ZDCF}.}
        \label{fig:dcfs}
\end{figure*}

\section{Timing Analysis}
\label{sec:time}

\subsection{Variability}
\label{sec:fvar}

X-ray variability of an AGN provides a powerful probe of the nearby regions of the central black hole.
Since Ark 120 has a `bare-type nucleus', the X-ray comes from the Compton cloud and is not intercepted
by any clouds such as BLR, NLR or molecular torus. Thus, the X-ray variability is originated from
the varying Compton cloud and the central accretion disc. To analyze the temporal variability in X-ray of 
Ark 120 in different energy bands, we have estimated different parameters for the duration 
of 2003 (MJD-52875) to 2018 (MJD-58118). The fractional variability $F_{var}$  
(\citep{Edelson1996}; \citep{Nandra1997}; \citep{Edelson2001}; \citep{Edelson2012}; 
\citep{Vaughan2003}; \citep{RP1997}) of lightcurves of $x_i$ count/s with finite 
measurement error $\sigma_i$ of length $N$ with a mean $\mu$ and standard 
deviation $\sigma$ is given by:

\begin{equation}
F_{var}=\sqrt{\frac{\sigma^2_{XS}}{\mu^2}}
\end{equation}
where, $\sigma^2_{XS}$ is excess variance (\citet{Nandra1997}; 
\citet{Edelson2002}), an estimator of the intrinsic source variance and is given by:
\begin{equation}
\sigma^2_{XS}=\sigma^2 - \frac{1}{N}\sum_{i=1}^{N} \sigma^2_{i}.
\end{equation}

The normalized excess variance is given by $\sigma^2_{NXS}=\sigma^2_{XS}/\mu^2$. 
The uncertainties in $\sigma^2_{NXS}$ and $F_{var}$ are taken from
\citet{Vaughan2003} and \citet{Edelson2012}. 

The X-ray variability of Ark 120 in different energy bands
($0.5 - 10.0$ keV; $0.2 - 2.0$ keV; $3.0 - 10.0$ keV) have demonstrated
different degrees of variabilities (Table \ref{tab:fvar}) while the time 
binsize is kept constant at $100$s. From XMM1, the lower energy 
($0.2-2.0$ keV) count rate was initially high ($X_{max}=21.95$) in 2003 
observation. Then, in 2013 (XMM2), it became half ($X_{max}=10.24$) from its initial value. 
In 2014 (XMM3), the count increased ($X_{max}=21.37$).
The fractional variability in this energy range increased from $0.016$ to $0.064$ from 2003 to 2014
observations. A similar trend is shown by $\sigma^2_{NXS}$ 
($0.006$ to $0.082$) in this energy band for each observation of XMM (Table \ref{tab:fvar}).
Like low energy part, the high energy ($3.0-10.0$ keV) follow the similar
type of trend for the count rate and fractional variability. The average value 
of $\sigma^2_{NXS}$ is $0.003$, with a range from $0.0014$ to $0.0056$.

We calculate the variability in $0.5-10.0$ keV range from the {\it Suzaku} data. We 
find higher variability $F_{var}=8.62\pm0.31$ in the 2007 {\it Suzaku} data as compared
to the previous XMM observations. The variability for XRT observations in $0.5-10.0$
keV range is shown in Table \ref{tab:fvar}. Due to the lack of data points, XRT1 observation 
yields an imaginary value of $F_{var}$, and is not shown in Table \ref{tab:fvar}. From 
the other observations of {\it Swift}/XRT, we observe high fractional variability 
($F_{var}$) from $0.14$ to $0.23$ with $<F_{var}>=18.22$. The average value of $x_{max}/x_{min}$ 
and $\sigma^2_{NXS}$ for these observations are $2.65$ and $0.045$ with a range from 
$2.16$ to $3.09$ and $0.027$ to $0.060$ respectively.  

\subsection{Delay Estimation}
\label{sec:delay_cal}
For temporal analysis of the long term archival data of Ark 120, we stress three epochs
of {\it XMM-Newton}, 2003, 2013, and 2014 out of which the latter two have high energy 
(3-80 keV) counterparts observed by {\it NuSTAR}. We have performed cross-correlation analysis 
using {\tt DCF} \citep{EK88} and $\zeta$-discrete cross-correlation function 
({\tt ZDCF}\footnote{{\tt ZDCF: }\burl{http://www.weizmann.ac.il/particle/tal/research-activities/software}}, 
\cite{A97}) for comparison. The likelihood is calculated using 12000 simulation
points in the {\tt ZDCF} code for the lightcurves obtained by {\it XMM-Newton}. The peak
error is calculated using the formula provided by \cite{gp87}. We have followed a similar 
procedure as in \cite{Ch20}. The time resolution of each light curve 
is $1000$s. The $0.2-2$ keV lightcurve obtained from 2003 data yields an acceptable $\chi^2_{red}<1.5$  
when fit with a straight line. However, data procured in 2013 and 2014 in a similar energy band have a
high residual and are not suitable for linear fitting. All three high energy lightcurves 
(3-10 keV) have $\chi^2_{red}<1.5$ when fitted with straight lines. We have carried out the delay 
estimation using the {\it XMM-Newton}/Epic-pn data to ensure the simultaneity in their procurements.  

\begin{table*}
\centering

    \caption{Parameters used in delay estimations are presented. The error in measurement of delay is
	considered as the larger between binsize and \protect{$\epsilon_{\tau}$}. \protect{$\epsilon^{d}_{\tau}$} and
	\protect{$\epsilon^{z}_{\tau}$} represents errors for {\tt DCF} and {\tt ZDCF} patterns.}    
    \label{tab:delay}

    \begin{small}
    \begin{tabular}{|c|c|c|c|c|c|c|}
    \hline
{\bf Id}&{\bf Epochs}&{\bf Bin size}&$\epsilon^{d}_{\tau}$&{\bf $\Delta\tau^{dcf}_{d}$}&$\epsilon^{z}_{\tau}$  &{\bfseries $\Delta\tau^{zdcf}_{d}$}\\
                             &{Year}      &  {(ks)}      &{(ks)}   &{(ks)}        &{(ks)}  &{(ks)}  \\
    \hline                                                                                                                        
    \textcolor{black}{XMM1}  &  2003      & 1            & 0.388   & $0.16 \pm 1$ &0.936   &   $-0.057 \pm 1$           \\
    \hline                                                                                                                        
    \textcolor{black}{XMM2}  &  2013      & 1            & 0.862   &$4.71 \pm 1$  &2.11    &   $6.76 \pm 2.11$          \\
    \hline                                                                                                                        
    \textcolor{black}{XMM3}  &  2014      & 1            & 0.622   &$-4.15\pm 1$  &1.54    &   $-4.56\pm 1.54$          \\
    \hline                                                                                                                        
    \textcolor{black}{XMM3}  &  2014      & 1            & $-do-$  &    $-do-$    &1.58    &   $-49.2\pm 1.58$          \\
    \hline
    \end{tabular}
    \end{small}
\end{table*}

The {\tt DCF} \citep{EK88}, performed using the lightcurves, have generated three 
distinct patterns. The 2003 data has produced $2.78 \pm 16.67$ minutes or 
$\sim 0.16$ ks delay. We have fitted the peak using a {\tt Gaussian}
model (dotted line in Fig. \ref{fig:dcfs}). Considering the error, no delay can be seen 
between two bands of X-ray. Similar delay pattern is also observed from {\tt ZDCF}, and the 
likelihood density also maximizes around zero. Likewise, we have performed {\tt Gaussian} 
fitting for 2013 data where a positive delay of $78.51 \pm 35.17$ minutes or $\sim 4.7$ 
ks has been seen between soft and hard X-ray photons using {\tt DCF}. But, the {\tt ZDCF}
peak maximizes around $112.68 \pm 35.22$ minutes or 6.7 ks and likelihood peak coincides 
with that (see Fig. \ref{fig:dcfs}). In 2014, the delay sign have switched, and we find a negative 
delay of $-69.19\pm 25.67$ minutes or $\sim -4.1$ ks between the soft and hard band from 
{\tt DCF} analysis. However, {\tt ZDCF} peaks maximize around two positions, $-76.19\pm 25.67$ ($-4.56 \pm 1.54$ ks)
and $-820.19\pm 26.46$ ($-49.2 \pm 1.58$ ks) minutes having peak values of 0.664 and 0.722 respectively. 
Between these two, the former coincides with the {\tt DCF} pattern (see, Table \ref{tab:delay}
for details). For all three cases, we find the peak values of {\tt ZDCF} patterns are lesser than the 
corresponding peak values obtained from {\tt DCF} patterns.  

\section{Discussions}
\label{sec:dis}

We have studied the central region of Ark 120 through X-ray (above 0.2 keV) 
using the data of {\it XMM, Suzaku, NuSTAR} and {\it Swift/XRT} in the 
period 2003 (MJD-52875) to 2018 (MJD-58118). As it is a bare type AGN,
the X-ray spectra mainly generated from the nearby region of the 
central engine. 

\subsection{Evolution of the Source: Primary Continuum}
\label{sec:pc}

The `bare-type AGN' Ark 120 was observed for a period of fifteen years, 2003 to 2018
using various X-ray satellites.
During these observations, the source has exhibited variabilities in both spectral 
and temporal domain. The luminosity of the source in the energy range of 2.0 to 10.0 keV  
varied within $\sim 10^{43.5}-10^{45.5}$ erg/s throughout these observations.
From the {\tt nthcomp} model, we report the variation of the spectral index (1.6<$\Gamma$<2.08)
where the harder spectra were observed after 2014. Following \cite{Vaughan2004}, we have
fitted the 2003 spectrum of Ark 120 with ({\tt nthcomp + Gaussian}) model. The fitted 
$\Gamma=1.90^{+0.01}_{-0.01}$ agrees with the spectral index previously observed 
(Table 4 of \cite{Vaughan2004}). Corresponding temperature of the Compton cloud is 
$kT_e=159.45^{+81.68}_{-81.69}$ keV. The ({\tt TCAF + Gaussian})
model provided a few previously unknown parameters like accretion rates, disc rate 
$\dot{m}_d=0.063\pm0.002$ and halo rate $\dot{m}_h=0.112\pm0.001$. This suggests that the
the source was initially halo dominated. This is normal for an AGN. 
The shock location or the size of the CENBOL ($X_s$), estimated from the fits, is 
$20.36\pm4.4~r_g$. The shock is found to be moderately strong with a compression ratio of $R=1.95\pm0.05$.   

The softest spectrum, having $\Gamma=2.08^{+0.03}_{-0.03}$ 
is seen during the {\it Suzaku} observation in 2007. It is to be noted that, \cite{Nardini2011} found 
the spectral index to be $\Gamma=2.03^{+0.01}_{-0.04}$ for the {\it Suzaku} data using blurred
reflection model. We have estimated the temperature of the Compton cloud to be 
$kT_e=124.65^{+35.54}_{-35.21}$ keV. This is the least of all temperatures  
obtained from all the observations. Using a single {\tt Gaussian}, we 
find the presence of a broad iron line ($6.38^{+0.052}_{-0.052}$) keV having an equivalent 
width of $EW=710^{+10}_{-10}$ eV. The derived optical depth is $\tau=0.726^{+0.008}_{-0.008}$. 
This suggests an optically thin Compton cloud. From the {\tt TCAF} fits, we find 
that the size of the Compton cloud has slightly increased to $X_s=21.44\pm4.9~r_g$ from the earlier 
observation. Corresponding disc rate, which enhances the soft seed photons, has 
increased to $\dot{m}_d=0.126$. Also, the halo rate has increased to
 $\dot{m}_h=0.191$. However, shock strength has decreased (see Table \ref{tab:4}). The 
drop in the $kT_e$ could be understood easily from TCAF, where the increase in disc 
rate leads to an enhanced cooling fraction. Thus, within the epochs of 2003 and 2007, 
the temperature of the Compton cloud was varied from 159.45 to 124.65 and as a result the spectrum 
softened.

\begin{figure*}
	\includegraphics[width=2.0\columnwidth]{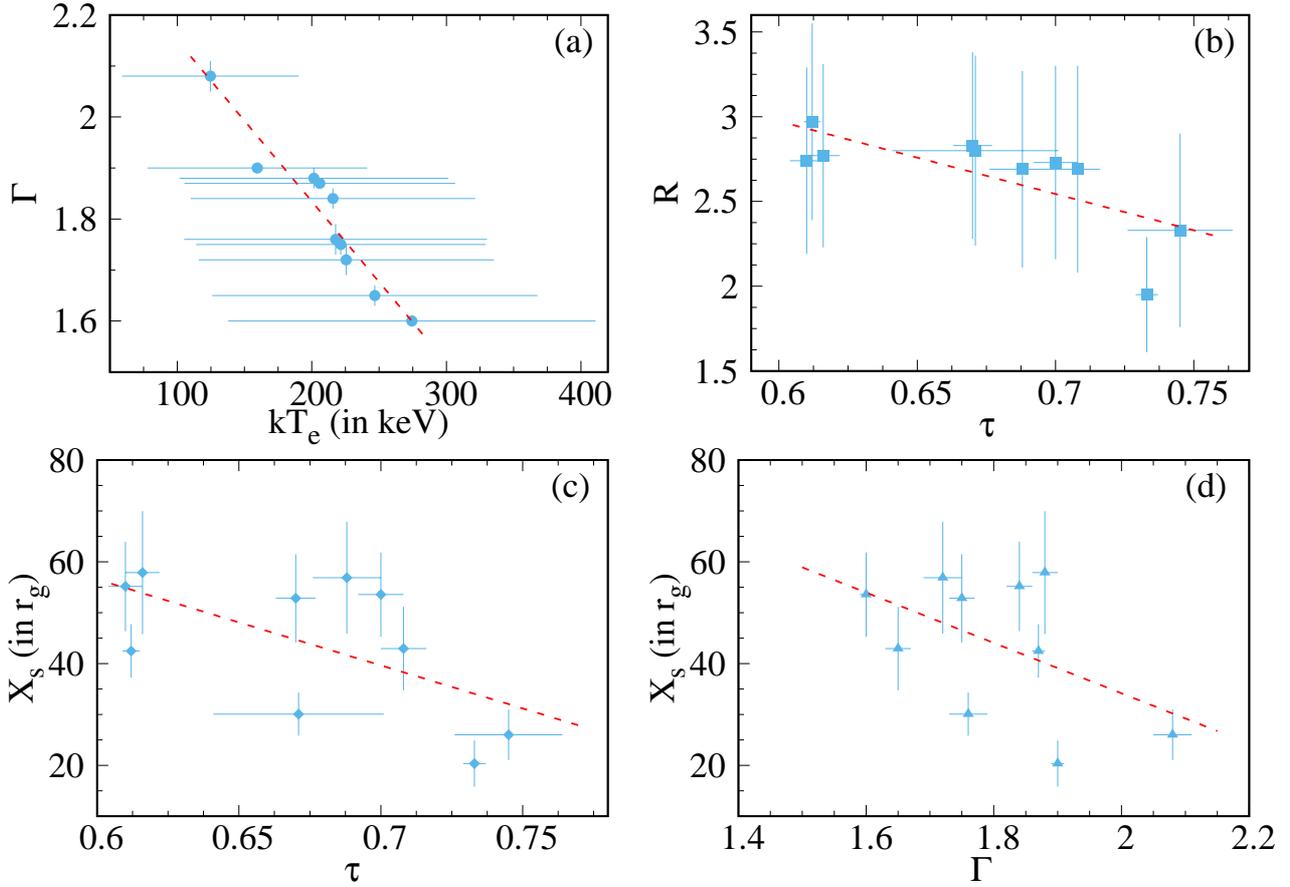}
	\caption{Correlation of fitted parameters are plotted. Fig. (a) represents the 
	 	correlation between $\Gamma$ vs $kT_e$ and the corresponding PCC is -0.95. It is 
		also noted that the $kT_e$ is losely bound with $\Gamma$. Fig. (b) is the correlation 
		between $\tau$ and R. The PCC for these parameters is -0.72. In Fig. (c) represents 
		the correlation between $\tau$ vs $X_s$ and the corresponding PCC is -0.45. Fig. (d) 
		provides the correlation of $\Gamma$ vs $X_s$ with PCC -0.53. }
	\label{fig:corr}
\end{figure*}

Later, in 2008, {\it Swift} observed the source where the spectrum hardened from the 
previous observation having $\Gamma=1.76^{+0.02}_{-0.08}$, $kT_e=217.72^{+105.6}_{-112.5}$ keV, 
and optical depth $\tau=0.671^{+0.030}_{-0.030}$. The iron line could not be 
detected from the XRT spectrum. Corresponding TCAF fitted parameters, such as the shock 
location $30.08~r_g$ and $R=2.80$ while $\dot{m}_d$ and $\dot{m}_h$ have changed to 
$0.064$ and $0.11$ respectively.

Significant variation of spectral properties is also noted during 2013 and 2014. 
The broad-band spectra (3-78) keV are fitted with ({\tt nthcomp + Gaussian}) 
having the spectral indices $1.75^{+0.01}_{-0.02}$ and $1.87^{+0.01}_{-0.01}$ and 
are in good agreement with parameters obtained by \cite{Porquet2018, Marinucci2019}. The optical 
depth is reduced from $\tau=0.670\pm0.074$ to $\tau=0.612\pm0.003$. The flux 
in 2-10 keV band has doubled within a year. The spectral softening 
could be explained by the drop of temperature of the Compton cloud. However, the 
decrease in the optical depth for March 2014 data with respect to 2013 has also
been seen from Monte-Carlo simulations \citep{Marinucci2019}. From TCAF fitting, 
we find a distinct variation of the flow parameters. The $\dot{m}_d$ changed from $0.068$ to $0.103$, 
$\dot{m}_h$ changed from $0.111$ to $0.126$, and $X_s$ changed from $52.83$ to $28.24$ within 2013 and 2014 
observations respectively. As the disc accretion rate increases, Compton cooling
increases, and this lead to the decrease in the $X_s$ which finally softens the 
spectrum. Considering TCAF, the lower optical depth for softer spectrum could be
explained by the weakening of the shock ($R=2.43$ as compared to $R=2.83$ in February 2013) 
for this observation. The stronger shock creates a distinct boundary between 
the halo and CENBOL region where the majority of the hard photons are produced.
However, for the weaker shock, the CENBOL boundary is less sharp and a fraction 
of inverse Comptonization could occur within the halo component. Thus, the 
effective optical depth of the medium could become lower even though the spectrum has 
softened.

Ark 120 has shown significant variabilities after February 2014 and is monitored by {\it 
Swift}. We have tabulated the spectral and temporal variabilities in Table \ref{tab:2}
and \ref{tab:4}. During September-October of 2014, we find that the spectral slope was $\Gamma=1.60^{+0.01}_{-0.02}$ 
and the corresponding temperature was $274.40\pm130.0$ keV, which was maximum within the duration
of our observation. From the TCAF fitting, we find $\dot{m}_d$ and $\dot{m}_h$ has 
changed to $0.068$ and $0.11$ respectively and the corresponding shock location has changed to $53.56\pm8.2~r_g$ and 
the shock strength has increased from $2.43$ to $2.73\pm0.5$  as observed during February 
2014. Later, in December 2014, the spectrum has softened with $\Gamma=1.84\pm0.02$ with the
temperature of Compton cloud $215.72\pm105.5$ keV. The corresponding shock has moved 
outward and observed at $55.16~r_g$ and $R=2.74$. Like previous observations, we 
see the halo rate and disc rates are fixed at $0.11$ and $0.068$, respectively. 

XRT4 and XRT5 observations were made starting from the end of December 2014 to March 
of 2015. During this time, the spectral indices are $1.74$ and $1.88$ respectively. 
The temperature and optical depths have also varied during this time. From TCAF fitting, 
we find the halo rate has decreased to $0.061$ in the XRT4 observation. However, the 
disc rate was constant. Again in XRT5 observation, halo rate has increased to $0.069$
while the disc rate remained the same. The shock location and the compression ratio remained 
constant (considering the errors) within this period. Thus, we can see that Ark 120 
exhibited spectral variability (see Fig. \ref{fig:evolve}) within $\sim$200 days 
(since September 2014-March 2015).

In XRT6, which was observed from December 2017 to January 2018, the spectrum of 
Ark 120 has hardened with respect to the earlier observations during January 2015.
The spectral index and temperature of Compton cloud are $1.65\pm0.02$ and 
$246.87\pm121$ keV respectively. From TCAF fitting, we find the disc and halo rates
have increased to $\dot{m}_d=0.081$ \& $\dot{m}_d=0.14$ 
respectively and the corresponding shock location settled 
at $42.95\pm8.0~r_g$. 

In Figure \ref{fig:corr}, we have plotted the correlations of a few spectral parameters. 
We find the spectral index and the temperature of the Compton cloud is anti-correlated
(Fig. \ref{fig:corr}a with Pearson Correlation Co-efficient (PCC) = -0.9542) for the long term observation. 
However, the values of $kT_e$ are poorly constrained with respect to spectral 
indices. This is a well-established relation and is generally found in case of 
AGNs and Galactic black holes. In Fig. \ref{fig:corr}b, we have presented the correlation 
between shock compression ratio and optical depth. We find $R-\tau$ produces anti-correlation
having PCC=-0.721. In general, stronger shocks are associated with the harder 
spectra where the optical depth is expected to be less \citep{CDC2016} and the 
corresponding shock location is also expected to be bigger. Keeping that argument, 
we also show the $X_s-\tau$ correlation where an anti-correlation (PCC=-0.457) has been 
observed from the long term data and presented in Fig. \ref{fig:corr}c. As a consequence, the
spectral softens due to the reduction of the shock location $X_s$ i.e., the size of 
the Compton cloud, we find a global trend of anti-correlation (PCC=-0.562) between 
$X_s-\Gamma$ (see Fig. \ref{fig:corr}) for Ark 120.

From the {\tt nthcomp} fitting, it can be found that the Compton cloud of the 
source was optically thin for the entire period of observation. 
Overall, we also noticed that the disc and halo rate is nearly constant
and they are $\sim0.07$ and $\sim0.11$ respectively for the majority of observations. 
But, we find a higher disc and halo rate in 2007 and 2014 observation. The shock location and 
the compression ratio have varied with time. The variation of these parameters is 
shown in Figure \ref{fig:evolve}. First, the shock location increases with time 
from 20 to 52 $r_g$ in the first $\sim 10$ years. Then the shock location 
falls  to 26.7$r_g$ within the next $\sim$ 13 months. Later, we find that the shock 
location again moves outward from 26.7 to 57.8 $r_g$ before moving inward 
again, and finally settling at 42.95$r_g$ in January 2018. The Compression 
ratio (R) also varies as the shock location ($X_s$). First, the compression ratio 
increased from 1.95 to 2.83 in $\sim10$ years. Then, the value of $R$ decreased 
to $1.67$ within next 1 year. After that, it increased to 2.73 within less than
six months and finally reached 2.69 at the end of January 2018. 

\subsection{Evolution of the Source: Delay patterns}
\label{sec:delay}

The Compton delay \citep{Payne1980,Sunyaev1980} for an electron cloud of size $\mathcal{R}$ having 
an optical depth $\tau$ and temperature $\theta_e=kT_e/m_ec^2$ can be described by, 
$$
t_{c} = \frac{\mathcal{R}}{c(1+\tau)}\frac{ln(E_h/E_{ss})}{ln[1+4\theta_e(1+4\theta_e)]},
$$
where, $c$ is the velocity of light, $E_h$ and $E_{ss}$ are the energy of hard photons and 
soft seed photons respectively. For AGNs having a central black hole mass of $1.5\times10^8$ 
\citep{Peterson2004}, the seed temperature of the photons remains in the 1-10 eV range. 
The maximum of the hard and soft energy band is considered to be $10$ keV and $1$ keV and the 
seed photon temperature is $E_{ss}=3$ eV. The light-crossing time for 1$r_g$ is $r_g/c=$1.5 ks 
for Ark 120. We calculated the delays for the combined parameters obtained from {\tt nthcomp} 
and {\tt TCAF} model. 

We have calculated the Compton delay for XMM1 observation where the size of the Compton cloud 
is $\sim 20~r_g$, optical depth $0.733$, and $\theta_e=0.311$. Substituting the values, we 
find $t^h_c=105.3$ ks and $t^s_c=75.3$ ks which produces a positive theoretical delay of 
$\Delta \tau = t^h_c-t^s_c=30$ ks. However, from the observed {\tt DCF} pattern, we fail to notice 
any such delay for this case. Here, we find light crossing delay ($\tau_{lc}$) of 30 ks for a $\sim 20~r_g$ 
Compton cloud. The observed {\it zero-delay} could be a combined result of $\tau_c$ and 
$\tau_{lc}$. In that case, it is to be noted that $\tau_{lc}$ becomes crucial in presence 
of a significant contribution of reflection component ($R_{ref}=1.96$, see Table \ref{tab:4}).

\begin{figure*}
	\centering {
	\includegraphics[height=0.73\columnwidth,width=1.0\columnwidth]{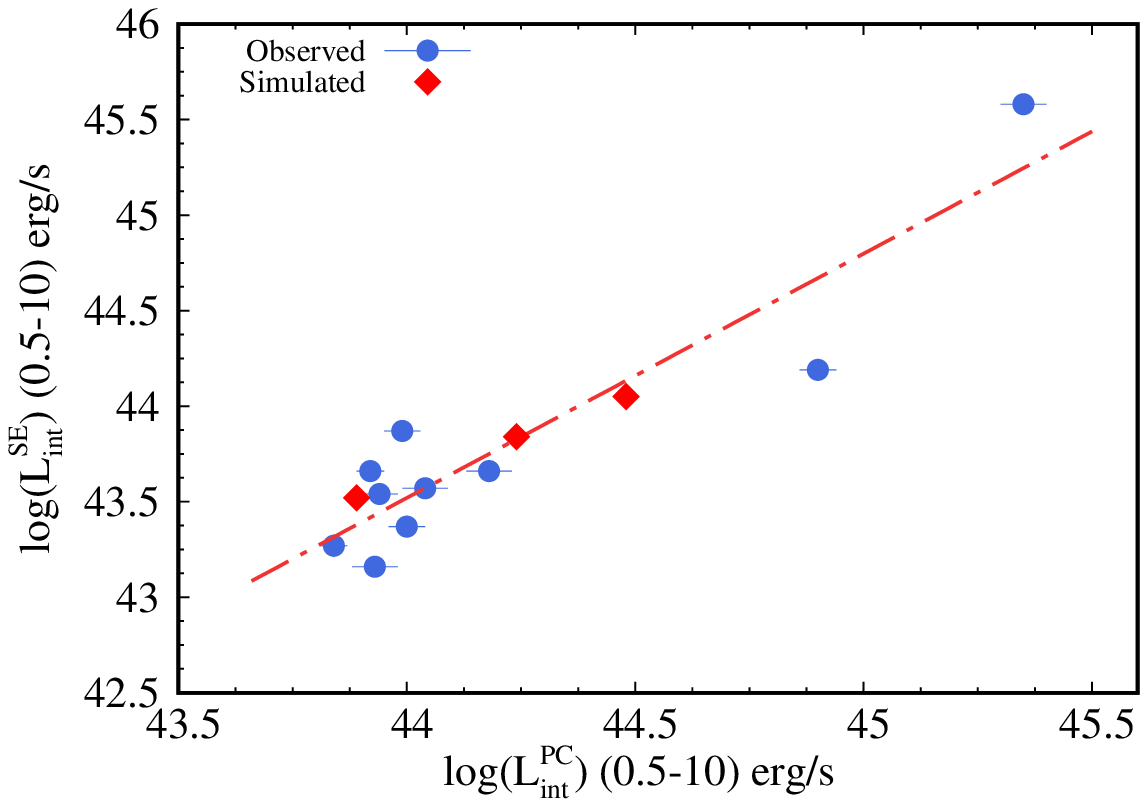}
        \includegraphics[width=1.0\columnwidth]{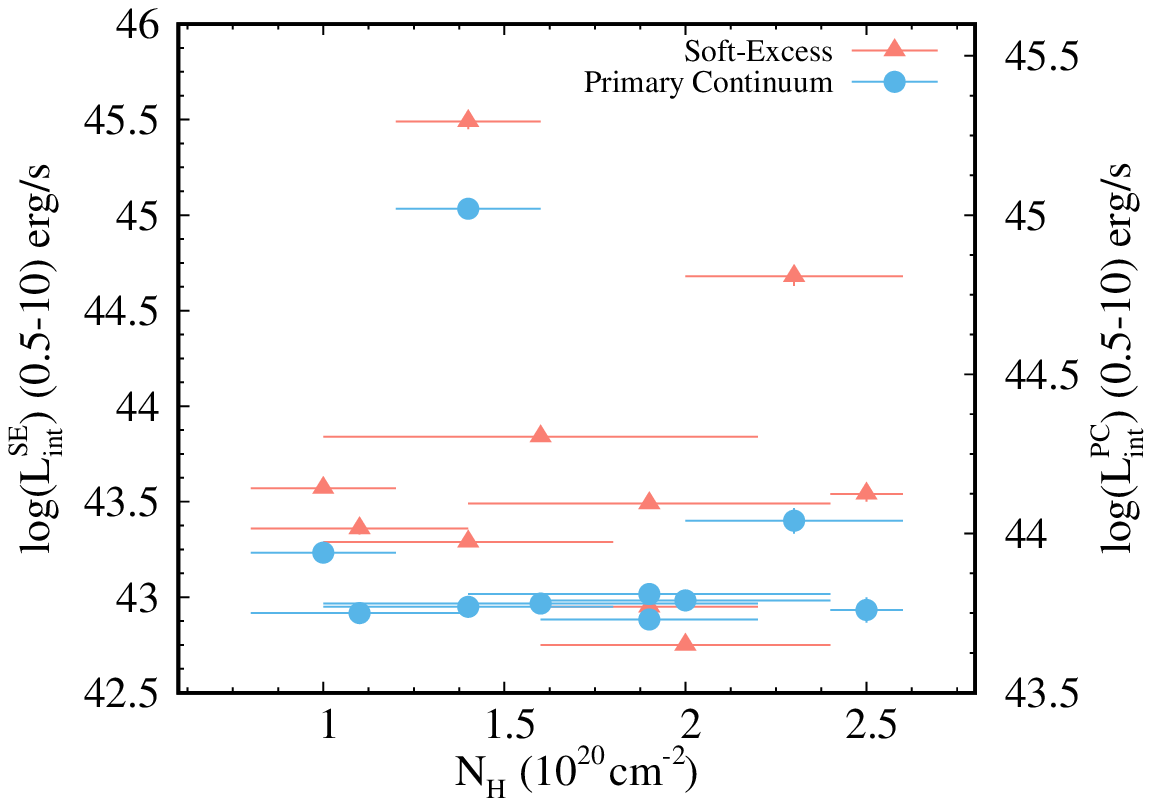}}
	\caption{Correlation of intrinsic luminosities of 0.5-10.0 keV 
	obtained using {\tt nthcomp}. {\it Left:} shows a correlation (PCC=0.92)
	between the observed intrinsic luminosities of primary continuum and soft-excess 
	(blue-circle). Monte-Carlo simulated luminosities for both energy ranges are presented with 
	red-diamond points. {\it Right:} No correlation between intrinsic luminosities and 
	$N_H$ from long term observations.}
	\label{fig:corr_param}
\end{figure*}

For the broadband observation (XMM2+N1), the size of the Compton cloud is $\mathcal{R}\sim 50$ 
$r_g$, having an optical depth of $0.67$ and temperature $\theta_e=0.434$. Combining all these, 
the maximum hard and soft energy delay which can be generated via Compton scatterings are 
$t^h_c=208$ ks and $t^s_c=148$ ks respectively. Thus, the maximum delay between hard and soft bands 
of X-ray can be $\Delta \tau = t^h_c-t^s_c=60$ ks. The light crossing delay is around 
$\tau_{lc}=75$ ks. The combined effects of $\Delta \tau$ and $\tau_{lc}$ should yield a 
negative delay of 15 ks. However, as discussed previously, $\tau_{lc}$ could dominate 
if reflection becomes dominating (here $R_{ref}=0.25$). Also, the size of the Compton 
cloud is much bigger than the what should be the `{\it transition radius}' 
(see, \cite{DC16,dpc18} for details) of an AGN having mass $1.5\times10^8~M_{\odot}$. Being an 
intermediate inclination angle source \citep{Nardini2011,Marinucci2019}, Comptonization
dominates the time delay when the size of the Compton cloud is bigger. The theoretical structure of 
Compton cloud is somewhat deviated from the sphere (see, \cite{CT95}) and the thermodynamical
fluctuations within the inhomogeneous Compton cloud (see, \cite{Ch17b}) contributes to 
the delay patterns. Considering this, the effect of light crossing delay would be much 
less and Comptonization could be considered as the core process, which generates $0.2-2$ 
keV photons during 2013 observations.

In a similar way, we calculate the Compton delay for broadband observation in 2014 (XMM3+N2). For 
that, the size of the Compton cloud $\mathcal{R}=28~r_g$, the optical depth is $\tau=0.612$, and $\theta_e=0.403$. We 
have obtained $t^h_c=123.4$ ks and $t^s_c=88.2$ ks which produces $\Delta \tau = t^h_c-t^s_c=35$ 
ks. Contrary to that, the observed delay is $-69.19\pm 16.67$. Clearly, 
the Comptonization may not be the dominating radiative process for this observation. From Table 
\ref{tab:4}, we see that the reflection co-efficient $R_{ref}=0.96$, which refers to a stronger
reflection. It is also to be noted that \cite{Lobban2018} found the X-ray to be leading the U-band
by $2.4\pm1.8$ days which they have explained with the light crossing delay. Considering 
the Compton cloud only, $\tau_{lc}$ becomes 42 ks, which is comparable to compensate for 
the positive lag obtained from Comptonization. In this particular case, the maximum possible
negative delay would be $\Delta \tau - \tau_{lc} \sim -7$ ks or -116 minutes. However, as the 
size of the Compton cloud has become bigger and $R_{ref}$ is much less than the XMM1 observation. 
Thus, the contribution from $\tau_{lc}$ could be less effective and we observe a negative 
delay much less than the maximum allowed delay.

Thus, along with the spectral variations, we find the delay patterns have varied over the
three epochs (2003, 2013, and 2014) in which {\it XMM-Newton} observed Ark 120. A significant
change in the delay pattern is observed within a year (2013-2014) where the positive delay
changed sign and becomes negative with a similar magnitude.

\subsection{Soft Excess}
\label{sec:soft}
The origin of ubiquitous {\it soft-excess} \citep{Arnaud1985,Singh1985,Brandt1993,Fabian2002,GD2004} 
remains debated. A plausible cause of soft-excess was given using 
reflection \cite{Sobolewska2007}. The multi-wavelength campaign of Mrk 509 
\citep{Mehdipour2011} revealed the correlation of soft-excess with the 
optical-UV part both in the spectral and temporal domains where they concluded 
that the soft-excess was generated due to Comptonization by a warm optically 
thick region surrounding the accretion disc. \cite{Done2012} proposed that
the high mass accretion rate of the disc could generate the soft-excess. 
For lower $L/L_{EDD}$, the energy dependent variability in the soft-excess 
part was found to be less in case of Narrow line Seyfert 1 galaxies. 
\cite{Lohfink2012} studied Seyfert 1 galaxy Fairfall 9 where the origin 
of the soft-excess component was found to be connected with source which generates 
the broad iron line. However, they implied that another source of Comptonization 
might be responsible for the formation of the soft-excess. 

\begin{figure*}
	\centering {
	\includegraphics[width=1.0\columnwidth]{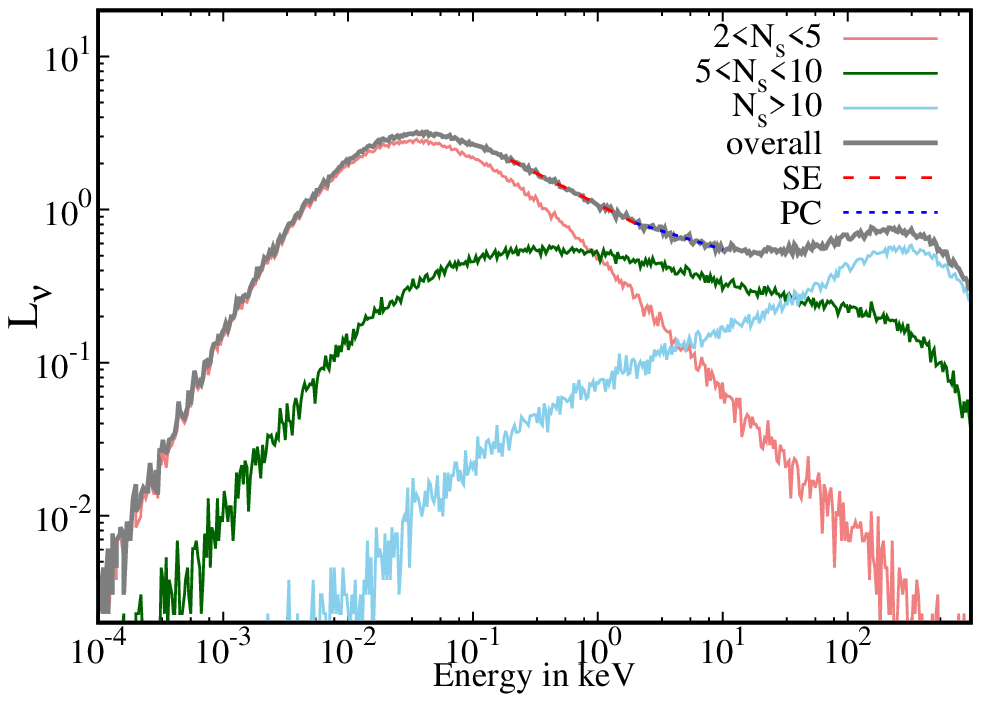}
	\includegraphics[width=1.0\columnwidth]{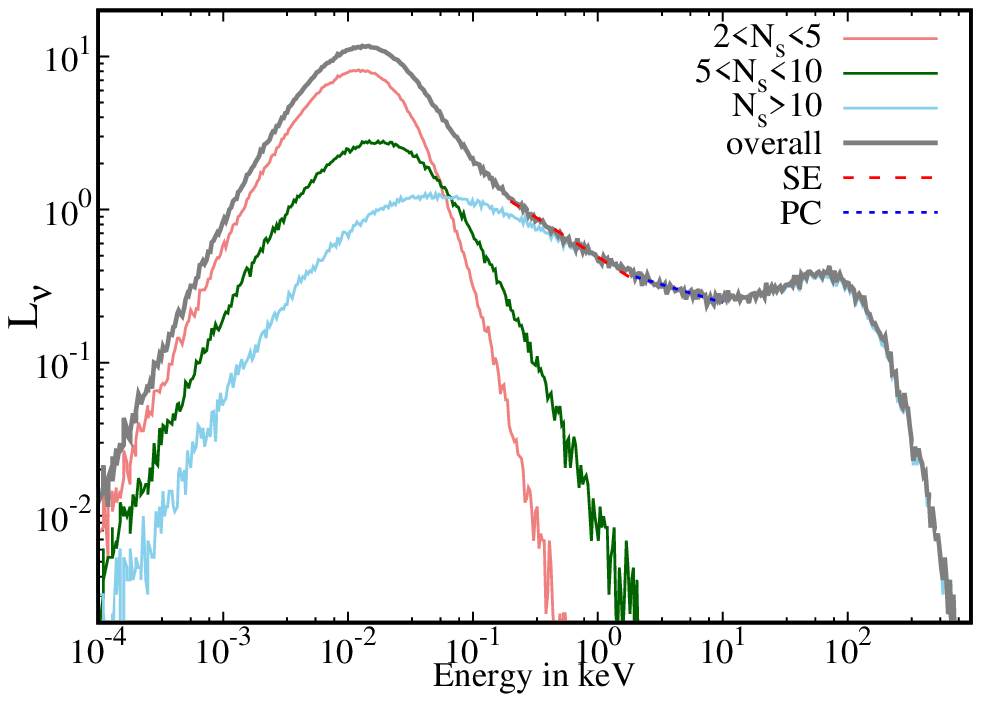}}
\caption{Monte-Carlo simulated spectra for Ark 120 are presented. We have considered 
	 $M_{BH} = 1.5\times10^8~M_{\odot}$ for Ark 120. Simulation boundary extends up to $100~r_g$. 
	 For {\it left panel} $\dot{m}_d$ = 0.06; $\dot{m}_h$ = 0.1, $X_s=60~r_g$, and maximum $kT_e=270$ keV. 
	 For {\it right panel} $\dot{m}_d$ = 0.1; $\dot{m}_h$ = 0.1, $X_s=40~r_g$, and maximum $kT_e=100$ keV. 
	 Notice the spectral contributions due to increasing number of scatterings. $L_{\nu}$ has been 
	 normalized with respect to the observed spectrum.} 
	\label{fig:mc}
\end{figure*}

A strong soft-excess present in the X-ray spectrum of Ark 120  was
reported by \cite{Brandt1993, Matt2014, Porquet2004}. This soft-excess is also free from the 
absorbers and was reported by \cite{Nardini2011}. As a first step, we investigate 
the spectral slopes and the relative contribution of the soft-excess from 2003 to 
2018 using the {\tt nthcomp+zGaussian+powerlaw} model and the results are presented in Table 
\ref{tab:SE}. Subsequently, we freeze the $\Gamma^{nth}$ obtained from {\tt nthcomp} while
fitting the soft excess below 3 keV. The $\Gamma^{pl}$ fits the soft-excess < 3 keV. For every observation, we find
a soft-excess steeper than the primary continuum (see, Table \ref{tab:SE}) which 
is a characteristic associated with the Narrow line Seyfert 1 galaxies.
Apart from the steeper power law, the variation of soft-excess luminosity and
spectral index can be observed from long term observations presented in
Table \ref{tab:SE}. We have calculated the intrinsic luminosities of {\tt nthcomp} 
and {\tt powerlaw} within the energy range 0.5 to 10.0 keV. In Fig. \ref{fig:corr_param}a,
we see a strong correlation (PCC=0.9227) between the intrinsic luminosities of soft-excess
($L^{SE}_{int}$) and primary continuum ($L^{PC}_{int}$). However, as a ``bare'' type 
AGN, Ark 120 has not shown any correlation (Fig. \ref{fig:corr_param}b) among the 
intrinsic luminosities and the line of sight hydrogen column density ($N_H$). 

While {\tt nthcomp} provides a good fit in the high energy range, we have used 
{\tt TCAF+zGaussian+pexrav} model (presented in Table \ref{tab:4}) in the entire 
range. We find that the {\tt TCAF} fits well in the range of $0.2-10$ and requires no other
additional model for the soft-excess part with the range of  $0.2-3$ keV. The fitted results
and residuals are presented in Fig. \ref{fig:2}. From the 
spectral fitting using {\tt TCAF}, one recognizes that the soft-excess could be 
originated from the photons which are rarely scattered in the Compton cloud. 
The surrounding halo will contribute to this energy band (0.2 - 2 keV). Also, some high
energy photons from the Compton cloud which could be reflected from the disc will appear 
in this energy range after losing their energy through reflection from the cold disc. We 
have performed Monte-Carlo simulations to show the spectral variations with $N_s$. This is 
briefly discussed in Sec. \ref{sec:mc}. 

\subsubsection{Simulated spectra}
\label{sec:mc}
Radiative and hydrodynamic origin of soft-excess has been investigated in
\cite{Fukumura2016} where they proposed that the shock heating near the ISCO could produce 
the soft-excess. The model reproduced the spectra of ``bare'' Seyfert 
1 galaxy, Ark 120. We have inspected the possibility of scattering dependent 
spectral contribution from the pre-shock and the post-shock regions \citep{CT95}. 
We extend the work of \cite{Gh11,Ch18} in case of AGNs considering Ark 120. 
Using the Total Variation Diminishing ({\tt TVD}) scheme \citep{Ryu97}, we inject
matter having a halo rate of $0.1$ from the outer boundary at $200~r_g$. TCAF
fitted parameters are used for the simulation setup and are mentioned in the 
Fig. \ref{fig:mc}. Considering the Keplerian disc in the equatorial 
plane ($z=0$), we construct the profile of the accretion disc 
following \cite{SS1973}. The Monte-Carlo simulation 
($0<r<100~r_g$) has followed the process provided by \cite{PSS83} and 
later extended by \cite{Gh09,Ch17a}. The simulations are performed using $10^7$ 
injected photons for each case. The emergent Comptonized spectra are plotted 
in Fig. \ref{fig:mc}. We show the variation of spectral components with respect
to the number of scatterings (see also \cite{Gh11}) within the region. 
From Fig. \ref{fig:mc}, we find that the spectra harden as the number 
of scatterings increase. The spectra of the primary component 
within the energy range 2.0 to 10.0 keV is dominated by the photons where the number 
of scatterings are $\ge 10$. However, the soft-excess, the red  
long-dashed line within 0.2-2 keV, is dominated by the contribution from 
photons which have suffered $\le10$ scatterings. A steeper spectral slope 
($\Gamma^{SE}$) for soft-excess is achieved with respect to the primary component 
($\Gamma^{PC}$) for both of the spectrum. This is similar to what has been observed 
for Ark 120 (Table \ref{tab:SE}). It is to be noted that, \cite{Boissay2016} 
studied the AGN 102 Sy1 and found that there is no link between the reflection 
and the soft excess. Instead, they indicated that the soft-excess could be related 
to the thermodynamical properties of Compton cloud and associated medium. 

\section{Conclusions}
\label{sec:con}
We have studied $\sim 15$ years of X-ray data of Ark 120.
We find the source varied considerably within that time span. This source was previously 
reported to be a `bare-type AGN' and we also find a similar nature 
of this source from the long term analysis. The X-ray count rate 
has increased by a factor of two in a few years, and it is not found 
to be related to the Hydrogen column density ($N_H$) since it is a 
`bare-type AGN'. Following are the major findings from our work.

\begin{enumerate}
\item[1.] The spectral slopes of the primary continuum ($\Gamma^{PC}$) 
and the soft-excess ($\Gamma^{SE}$) are not constant throughout our observational 
time span. $\Gamma^{PC}$ has varied between 1.60 and 2.08 whereas $\Gamma^{SE}$ 
between 2.52 and 4.23 from 2003 to 2018.  

\item[2.] The variation is reflected in fitted parameters of {\tt TCAF},
namely, the accretion rates and properties of the Compton cloud.
From the spectral fitting using {\tt TCAF}, we find that the  
disc rate ($\dot{m}_d$) and the halo rate ($\dot{m}_h$) have varied between
$0.061$ and $0.126$ and between $0.108$ and $0.191$ respectively. The shock location
($X_s$) or the size of the Compton cloud and compression ratio ($R$) 
vary correspondingly. $X_s$ varies between $20.36$ and $57.87$, whereas $R$ varies 
between $1.66$ and $2.73$.

\item[3.] We focussed on the simultaneous observations in low ($0.2-2.0$ keV)
and high ($3.0-10.0$ keV) energy X-ray band from {\it XMM-Newton} to 
calculate the time delay between them. We find that in 
XMM1 observation, there is no delay between the low and high energy band, while
a positive delay of $4.71\pm1$ ks is detected in XMM2 observation 
and a negative delay of $4.15\pm1$ ks is seen in XMM3 observation. 
A correlated variability among the optical, UV, and X-ray bands have
already been reported \cite{Lobban2020}. Also, \citep{DC16,Ch17b} reported
in a different context that the X-ray lag has a strong dependency on the 
geometric structure of the Comptonization region and orientation of the Keplerian disc.
The net delay is a resultant effect of different physical mechanisms, e.g., 
Comptonization, reflection, focusing, and jet/outflow emission \citep{Ch19,pa19}. 
For the lower inclination and radio-quiet nature of Ark 120, the positive delay 
could be attributed to the Compton delay while reflection and light-crossing delay 
could contribute to the negative delay.

\item[4.] From the analysis of the long term data, we report that the luminosity 
is independent of Hydrogen column density ($N_H$). This is expected as the source 
has a negligible line-of-sight hydrogen column density ($N_H<5\times10^{20}$). The luminosity 
of the primary continuum is highly correlated (PCC$\sim 0.92$) with the soft excess emission. From 
{\tt TCAF} fitting and Monte-Carlo simulations using TCAF flow configurations,
we show that the soft-excess spectral slope 
($\Gamma^{SE}$) is the result of a fewer Compton scatterings in the Compton cloud 
and the primary continuum ($\Gamma^{PC}$) is the result of the higher number 
of Compton scatterings. Corresponding intrinsic luminosities obtained from simulations 
corroborate  with the observed pattern. 

\end{enumerate}

\section*{Acknowledgements}
PN acknowledges CSIR fellowship for this work. AC acknowledges Post-doctoral 
fellowship of S. N. Bose National Centre for Basic Sciences, Kolkata India, 
funded by Department of Science and Technology (DST), India. BGD acknowledges 
Inter-University Centre for Astronomy and Astrophysics (IUCAA) for the Visiting 
Associateship Programme. This research has made use of data and/or 
software provided by the High Energy Astrophysics Science Archive Research 
Center (HEASARC), which is a service of the Astrophysics Science Division 
at NASA/GSFC and the High Energy Astrophysics Division of the Smithsonian 
Astrophysical Observatory. This work has made use of data obtained from the 
{\it Suzaku}, a collaborative mission between the space agencies of Japan 
(JAXA) and the USA (NASA). This work made use of data supplied by the UK 
Swift Science Data Centre at the University of Leicester. This work has 
made use of data obtained from the {\it NuSTAR} mission, a project led 
by Caltech, funded by NASA and managed by NASA/JPL, and has utilized the NuSTARDAS
software package, jointly developed by the ASDC, Italy and Caltech, USA. This
research has made use of the NASA/IPAC Extragalactic Database (NED) which is 
operated by the Jet Propulsion Laboratory, California Institute of Technology, 
under contract with the National Aeronautics and Space Administration. 
This research has made use of the SIMBAD database, operated at CDS, Strasbourg,
France.

\section*{Data Availability}
We have used archival data for our analysis in this manuscript. All the 
softwares used in this manuscript are publicly available. Appropriate links are given
in the manuscript.

\bsp    
\label{lastpage}
\end{document}